\documentclass[prd,aps,floats,letterpaper,floatfix,superscriptaddress, %nofootinbib,
twocolumn,preprintnumbers]{revtex4}
\usepackage{amsmath}
\usepackage{hyperref}
\usepackage{color}
\usepackage[mathenv]{}
\usepackage{graphicx}

\newcommand{\subbs}[1]{{#1}_{\mbox{\tiny BS}}}
\newcommand{\subm}[1]{{#1}_{\mbox{\tiny M}}}
\newcommand{\subl}[1]{{#1}_{\mbox{\tiny L}}}
\newcommand{\tinysub}[2]{{#1}_{\mbox{\tiny #2}}}

\newcommand{\beq}{\begin{equation}}
\newcommand{\bea}{\begin{eqnarray}}
\newcommand{\eeq}{\end{equation}}
\newcommand{\eea}{\end{eqnarray}}

\newcommand{\cvec}[2]{\begin{array}{c} #1 \\ #2 \end{array}}
\newcommand{\rvec}[2]{\begin{array}{cc} #1 & #2 \end{array}}
\newcommand{\comment}[1]{{\bf [commented out]}}
\begin{document}
\title{Mathematical framework for simulation of quantum fields in complex interferometers using the two-photon formalism}

\author{Thomas Corbitt}
\affiliation{LIGO Laboratory, Massachusetts Institute of
Technology, Cambridge, MA 02139}

\author{Yanbei Chen} \affiliation{Theoretical Astrophysics, California
Institute of Technology, Pasadena, CA 91125}
\affiliation{Max-Planck-Institut f\"ur Gravitationsphysik, Am M\"uhlenberg 1, 14476 Golm, Germany}

\author{Nergis Mavalvala}
\affiliation{LIGO Laboratory, Massachusetts Institute of
Technology, Cambridge, MA 02139}

\begin{abstract}
\noindent We present a mathematical framework for simulation of
optical fields in complex gravitational-wave interferometers. The
simulation framework uses the two-photon formalism for optical
fields and includes radiation pressure effects, an important
addition required for simulating signal and noise fields in
next-generation interferometers with high circulating power. We
present a comparison of results from the simulation with
analytical calculation and show that accurate agreement is
achieved.
\end{abstract}

\definecolor{purple}{rgb}{0.6,0,1}
\preprint{\large \color{purple}{LIGO-P030070-00-R}}
\preprint{\large \color{purple}{AEI-2005-006}}

\maketitle

\section{Introduction}
Next-generation gravitational-wave (GW) interferometers, such as
those planned for Advanced LIGO~\cite{pfspie}, are designed to
have a fifteen-fold improvement in sensitivity over present-day
detectors~\cite{LIGOI}. Among the techniques planned to achieve
this improved sensitivity is an increase in the input laser power.
The higher laser power reduces the shot noise limit at frequencies
above $\sim$100~Hz, as intended, but has the deleterious effect of
increasing the radiation-pressure noise at lower
frequencies.  Consequently,
advanced detector sensitivity at almost all frequencies in the
detection band is expected to be limited by {\it quantum} noise.
Qualitatively speaking, shot noise and radiation-pressure
noise correspond to measurement noise and back action noise in quantum
measurement theory --- together they often impose the Standard Quantum
Limit (SQL) to measurement accuracy~\cite{BK92}. A correct modeling of the
quantum noise of a GW interferometer should take into account {\it
  correlations} between the two types of noises, which may allow
sub-SQL sensitivities to be achieved~\cite{BK92,KLMTV,BC2}.

The need for optical field simulation for gravitational-wave
interferometer design has been addressed in the past with a
variety of simulation tools, both in the frequency domain (e.g.,
{\it twiddle}~\cite{twiddle} and {\it
  finesse}~\cite{finesse}) and in the time
domain (e.g., the {\it LIGO end-to-end simulation
program}~\cite{e2e}). Although time-domain simulations can study
issues associated with large mirror displacements and non-linear
effects, e.g., the lock acquisition of the interferometer, they
are computationally costly; in addition, full time-domain
simulations are also less straightforward to quantize. In order to
study the performance of gravitational-wave detectors, it suffices
to stay in the linear regime near the operation point. For such a
linear problem, frequency-domain simulations are dramatically
simpler than time-domain ones; it is straightforward to obtain
frequency-domain transfer functions, and therefore noise spectra.
In addition, since the system is linear, the propagation of
quantum Heisenberg operators are identical to those of classical
field amplitudes, therefore it suffices to build an essentially
classical propagator.

In low-power situations where radiation-pressure-induced mirror
motion is negligible and no non-linear optical elements (e.g.,
squeezers) are used, when linearizing over mirror displacements,
propagation of electromagnetic fields at different frequencies are
independent, and therefore the transfer functions can be
established for each different frequency separately. One only
needs to take into account that, for the inputs to this linear
system: (i) mirror motion (with frequency $\Omega$) creates phase
modulation of the carrier, which is equivalent to generating two
equally spaced sidebands on the carrier frequency (at
$\omega\pm\Omega$, where $\omega$ is the carrier frequency and we
denote $\omega + \Omega$ and $\omega - \Omega$ as the upper and
lower sidebands, respectively) with opposite amplitudes, and that
(ii) laser noise can usually be decomposed into amplitude noise
and phase noise, with the former contributing equally to the upper
and lower sidebands, and the latter oppositely. These
considerations have been the conceptual foundations of previous
frequency-domain simulation programs.

For high-power interferometers, the above strategy will have to be
modified: the radiation-pressure forces acting on the mirrors, at
frequency $\Omega$, depend on both upper and lower sideband fields;
the induced mirror motion will again contribute to both sidebands ---
this makes it necessary to propagate pairs of upper and lower
sidebands simultaneously. The mathematical formalism most convenient
for this problem, at least in the case of only one carrier frequency,
is the Caves-Schumaker {\it two-photon
  formalism}~\cite{CavesSchumaker,SchumakerCaves}. In this paper, we
adopt this formalism and present a mathematical framework for
calculating the propagation of fields in an arbitrary optical
system that includes the dynamical response of the mirrors to the
light field. Namely, we divide complex interferometers into
inter-connected elementary subsystems, and provide a general
procedure for building a set of linear equations for all optical
fields propagating between these systems -- based on each
individual system's {\it input-output relation}, i.e.,
transformation matrices relating output fields to input ones and
the incoming GW. We also describe the way in which these
subsystems are connected to each other. Solving these equations
will provide us with the optical fields, in terms of vacuum
fluctuations entering the system from open ports, laser noise, and
incoming GWs. While this mathematical framework, and the resulting
numerical simulation tool, were developed to model quantum
correlation effects in gravitational-wave interferometers, the
method is general and can be used in any system where optical
fields couple to mechanical oscillation modes.

The paper is organized as follows: In Sec.~\ref{sect:framework} we
introduce the mathematical framework for the simulation, and
illustrate it with a simple example; in Sec.~\ref{sect:matrices}
we provide input-output relations of basic optical elements that
may be present in a laser interferometer, ignoring
radiation-pressure effects and the presence of gravitational waves
--- by re-formatting well-known results in optics; in
Sec.~\ref{sect:radpres}, we take radiation-pressure-induced mirror
motion into account, and provide input-output relations for
movable mirrors and beamsplitters (up to linear order in mirror
motion), which have not been obtained before in the most general
form; in Sec.~\ref{sect:gwsignal}, we take into account the
presence of GWs by introducing modulation of cavity lengths, and
treat the corresponding effect on light propagation up to linear
order in $L/\lambda_{\rm GW}$ (with $L$ the length of the
interferometer). In Sec.~\ref{sect:ifoexample} the formulation is
applied to a novel interferometer designed to extract squeezed
vacuum states that are created by a strong opto-mechanical
coupling; and, finally, conclusions are summarized in
Section~\ref{sect:conclusions}.

\section{Mathematical framework}
\label{sect:framework}

\subsection{General Prescription}

% Figure 1
\begin{figure}[t]
\includegraphics[width=0.45\textwidth]{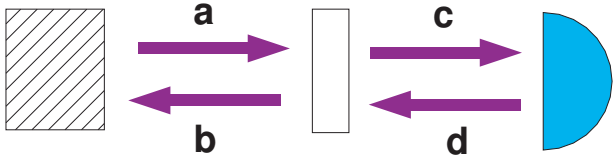}
\caption[Figure] {\label{fig:fig1} A sample configuration is
shown. A beam block is connected to a mirror, which is in turn
connected to a detector. Input fields incident on the mirror,
$\mathbf{a}$ and $\mathbf{d}$, are related to the output fields,
$\mathbf{b}$ and $\mathbf{c}$, by matrix operators derived in
Sections ~\ref{sect:framework}, \ref{sect:matrices} and
\ref{sect:radpres}.}
\end{figure}

As mentioned above, the presence of opto-mechanical coupling
dictates that we propagate the upper and lower sidebands
simultaneously, which means that for each frequency $\Omega$, we
will have to work with the two-dimensional linear space spanned by
the upper [$a(\omega+\Omega)$] and lower [$a(\omega-\Omega)$]
sidebands~\footnote{Strictly speaking, we have to consider the
four-dimensional linear space spanned by $a(\omega+\Omega)$,
$a(\omega-\Omega)$ and their Hermitian conjugates,
$a^\dagger(\omega+\Omega)$, $a^\dagger(\omega-\Omega)$. However,
the fact that the sideband fields are real functions in the time
domain will limit us to a two-dimensional subspace.}. Within the
{\it two-photon formalism}, developed by Schumaker and
Caves~\cite{CavesSchumaker,SchumakerCaves}, and outlined in
Appendix~\ref{app:quad} below, instead of $a(\omega\pm\Omega)$,
the two {\it quadrature fields} $a_{1,2}(\Omega)$ are chosen as
the basis vectors. For simplicity of notation, we generally denote
\begin{equation}
\mathbf{a} \equiv \left( \begin{array}{c} a_1 \\ a_2 \end{array}\right)
\end{equation}
and suppress the dependence of $\mathbf{a}$ on $\Omega$.

We consider optomechanical systems formed by the following
elementary subsystems: movable mirrors, beamsplitters and free
space propagators. We will also include a ``linear squeezer'',
which turns an ordinary vacuum state into a two-mode squeezed
field with arbitrary squeeze factor and squeeze angle. Auxiliary
to these optical elements, we introduce the beam block and the
photodetector to deal with open ports which are either left
undetected and those detected with unit quantum efficiency; we
also introduce the laser as an optical element, which injects
monochromatic carrier light and laser noise into the
interferometer. Quadrature optical fields undergo linear
transformations when propagating through such elementary systems,
and quadrature fields with different $\Omega$'s propagate
independently from each other. These linear transformations are
described mathematically by the {\it input-output relation},
namely a set of equations relating the output fields to the input
ones, including vacuum fluctuations, the carrier laser and laser
fields, as well as to incoming GWs. We provide these input-output
relations in Secs.~\ref{sect:matrices}--\ref{sect:gwsignal}.

However, we note that propagation of sideband quadratures
($\Omega\neq 0$), although independent from each other, all depend
on the propagation of the carrier quadratures ($\Omega =0$), i.e.,
the amplitude and phase of the carrier incident on each subsystem.
Fortunately, the propagation of the carrier is {\it not} affected
by that of the sidebands, and can be carried out independently at
the beginning. This said, we begin to formulate our general method
of simulation.

We build the following system of linear equations (for each
sideband frequency $\Omega$)
\begin{equation}
\label{general_equation}
\left[
\begin{array}{ccc}
\mathcal{M}_{11} &\cdots& \mathcal{M}_{1N} \\
\vdots &\cdots & \vdots \\
\mathcal{M}_{N1} & \cdots & \mathcal{M}_{NN}
\end{array}
 \right]
\left[
\begin{array}{c}
\mathbf{a}^{(1)} \\
\vdots \\
\mathbf{a}^{({N})}
\end{array}
\right]
=
\left[\begin{array}{c}
\mathbf{u}^{(1)} \\
\vdots \\
\mathbf{u}^{({N})}
\end{array}
\right]\,,
\end{equation}
where $\mathbf{a}^{(i)}$, $i=1,\ldots,N$ are the $N$ quadrature
fields (each of them a two-dimensional vector) propagating in
every part of the system, $\mathbf{u}^{(i)}$, $i=1,\ldots N$ are
$N$ generalized input quadrature fields (each of them again a
two-dimensional vector). The $\mathcal{M}_{ij}$, $i,j=1,\ldots N$
are $2\times2$ matrices which depend on the details of the optical
system, and the $\mathbf{u}^{(i)}$ can be written schematically as
\begin{equation}
\label{general_u}
\mathbf{u}^{(i)} = \mathbf{v}^{(i)}+  \mathbf{l}^{(i)} + \mathbf{H}^{(i)}h\,,
\end{equation}
where $\mathbf{v}^{(i)}$ arises from vacuum fluctuations entering
from the detection port or other lossy ports
(Secs.~\ref{subsect:blockpd}, \ref{sect:matrices} and
\ref{sect:radpres}), $\mathbf{l}^{(i)}$ from the laser
(Sec.~\ref{subsect:blockpd}), and $\mathbf{H}^{(i)}h$ from
GW-induced phase modulation, with $h$ the GW amplitude
(Sec.~\ref{sect:gwsignal}); depending on the location of this
generalized input field, some or all of the above three
contributions could also be zero. Henceforth in the paper, we
shall consider each pair of quadrature fields as one object.
Inverting the matrix $\mathcal{M}_{ij}$ will give
$\mathbf{a}^{(i)}$ in terms of $\mathbf{u}^{(i)}$, and hence all
of the necessary transfer functions.

Now let us provide a universal prescription for constructing
Eq.~\eqref{general_equation}, suitable for modelling generic systems.
We break this procedure into two steps:
\begin{enumerate}
\item
Suppose we have $n$ elementary subsystems mentioned above, with the
$k^{\rm th}$ subsystem having $p_k$ ports. The entire system will then
have $P\equiv \sum_{k=1}^n p_k$ ports. Because we formally include
beam blocks and photodetectors as subsystems, none of our ports will
be formally open, i.e., left unconnected to some other port. This
means that we have $P/2$
pairs of connections. For each pair of connections, we have two
fields, one propagating in each direction. This means we have a total
of $P$ fields (each in turn has two quadrature components).
\item
For each system $k$, with $p_k$ ports, we also have $p_k$ input
fields and $p_k$ output fields, and therefore the input-output
relation will provide us $p_k$ equations. All subsystems together
will then provide us with $P$ equations (each with two
components), exactly the number needed.
\end{enumerate}

\subsection{Example with the input-output relation of beam blocks, photodetectors and lasers}
\label{subsect:blockpd}

Next we illustrate the generic construction procedure with a
simple example, which also clarifies the formal roles of beam
blocks, photodetectors, and lasers. We first propagate fields
between three basic elements of an optical train: a beam block, a
partially reflecting mirror, and a photodetector. Referring to
Fig.~\ref{fig:fig1}, the beam block is connected to the mirror,
which is in turn connected to a detector. For simplicity, we
assume that the mirror is lossless and fixed in position.

As a first step, we identify the fields in consideration. The beam
block and the photodetector are 1-port systems, the mirror is a
2-port system; we have a total of 4 ports, and $4/2=2$
connections. There are two fields associated with each connection;
we label them $\mathbf{a}$, $\mathbf{b}$, and $\mathbf{c}$,
$\mathbf{d}$, respectively, as done in Fig.~\ref{fig:fig1}. Note
that each field in turn has two quadrature components, so the
system is 8-dimensional, and we need 8 scalar equations.

Now we have to provide the input-output relations for each object.
For the mirror with amplitude reflectivity $\rho$ and
transmissivity $\tau$, and neglecting radiation pressure effects,
we have
\begin{equation}
\label{eq:mirror}
\left(
\begin {array}{ccc}
\textbf{b}\\
\noalign{\medskip}
\textbf{c}
\end {array}\right) =
\left(\begin {array}{ccc}
-\rho&\tau\\
\noalign{\medskip}
\tau&\rho
\end {array}\right)
\left(\begin {array}{ccc}
\textbf{a}\\
\noalign{\medskip}
\textbf{d}
\end {array}\right)\equiv \textbf{M}_{\rm Mir}\left(
\begin {array}{ccc}
\textbf{a}\\
\noalign{\medskip}
\textbf{d}
\end {array}
\right).
\end{equation}
Note that Eq.~\eqref{eq:mirror} contains 4 scalar equations, and
that $\rho$ and $\tau$ are really $2\times 2$ scalar matricies,
$\rho \mathbf{I}$, and $\tau\mathbf{I}$ (this is true because our
mirror does not mix quadratures) --- we have suppressed the
identity matrix $\mathbf{I}$ for simplicity. To comply with the
format of Eq.~\eqref{general_equation}, we write
\begin{equation}
\label{mirror:ala:general}
\left(
\begin{array}{cccc}
-\rho & -1 & 0 & \tau \\
\tau & 0 & -1 & \rho
\end{array}
\right)
\left(\begin {array}{cccc}
\textbf{a}\\
\noalign{\medskip}
\textbf{b}\\
\noalign{\medskip}
\textbf{c}\\
\noalign{\medskip}
\textbf{d}
\end {array}\right)
=
\left(
\begin{array}{cc}
0 \\ 0
\end{array}
\right)\,.
\end{equation}
For the beam block and the photodetector, they really are placeholders
for physically open ports. Their input-output relation is
simply that the output fields from them are vacuum fluctuations
(independent from the input fields):
\begin{equation}
\label{eq:vac:d}
\mathbf{a} = \mathbf{v}^{(1)}\,,\qquad
\mathbf{d} = \mathbf{v}^{(2)}\,,
\end{equation}
Here we assume implicitly that the photodetector is detecting the
field $\mathbf{c}$ with unit quantum efficiency. In order to model
imperfect photodetectors, we could add a mirror with zero reflectivity
and non-zero loss in front of the ideal photodetector.

Combining Eqs.~\eqref{mirror:ala:general} and \eqref{eq:vac:d}, we
have
\begin{equation}
\label{eq:matrix_example}
\underbrace{\left(\begin {array}{cccc}
-1&0&0&0\\
\noalign{\medskip}
-\rho&-1&0&\tau\\
\noalign{\medskip}
\tau&0&-1&\rho\\
\noalign{\medskip}
0&0&0&-1
\end {array}\right)}_{\displaystyle{\mathcal{M}}}
\left(\begin {array}{cccc}
\textbf{a}\\
\noalign{\medskip}
\textbf{b}\\
\noalign{\medskip}
\textbf{c}\\
\noalign{\medskip}
\textbf{d}
\end {array}\right) =
\left(\begin {array}{ccc}
-\textbf{v}^{(1)}\\
\noalign{\medskip}
0\\
\noalign{\medskip}
0\\
\noalign{\medskip}
-\textbf{v}^{(2)}
\end {array}\right)\,,
\end{equation}
which are the 8 scalar equations we need. Inverting $\mathcal{M}$
will give us each of the propagating fields in terms of the input
vacuum fields.

Now suppose the beam block is replaced by a laser source, coupled
to the spatial mode of $\mathbf{a}$ field, then we only need to
replace the vacuum field $\mathbf{v}^{(1)}$ in
Eqs.~\eqref{eq:vac:d} and \eqref{eq:matrix_example} by the laser
field, $\mathbf{l}^{(1)}$: at $\Omega=0$, carrier quadratures,
while at $\Omega\neq 0$, it gives the laser noises.

Here we note that all diagonal elements of $\mathcal{M}$ are equal
to $-1$ --- this is in fact not a coincidence, but a universal
feature of our construction procedure. In order to understand
this, we need to realize that every field $\mathbf{a}^{(k)}$ is
the output field of exactly one subsystem. In the input-output
relation of that unique subsystem, there is exactly one line that
relates $\mathbf{a}^{(k)}$ to the input fields of this subsystem,
which reads:
\begin{equation}
\mathbf{a}^{(k)} = [\mbox{terms not involving $\mathbf{a}^{(k)}$}]\,.
\end{equation}
This equation corresponds to, after moving $\mathbf{a}^{(k)}$ to the
right-hand side of the equation, moving any non-$\mathbf{a}^{(j)}$,
$j=1,\ldots,N$ terms to the left-hand side, and swapping left and
right,
\begin{equation}
\big(
\cdots \;  \underbrace{-1}_{\mbox{$k^{\rm th}$ column}}\;   \cdots
\big)
\left(
\begin{array}{c}
\mathbf{a}^{(1)}\\
\vdots \\
\mathbf{a}^{(k)} \\
\vdots
\\
\mathbf{a}^{(N)}
\end{array}
\right)
=\ldots\,.
\end{equation}
It is obvious that the lines of equation found by this way for
different $\mathbf{a}^{(k)}$'s will be different. As a
consequence, we can arrange to have the line corresponding to
$\mathbf{a}^{(k)}$ appear on the $k^{\rm th}$ row of
$\mathcal{M}$, and thus have all its diagonal elements equal to
$-1$.

\section{Matrices for static optical elements}
\label{sect:matrices}

In this section, we derive the matrices for some standard objects
used in simulating quantum noise in a gravitational-wave
interferometer. Here we neglect radiation pressure effects and the
presence of gravitational waves (they will be dealt with in
Secs.~\ref{sect:radpres} and \ref{sect:gwsignal}, respectively).
As a consequence, our derivation only involves some re-formatting
of previously well-known results.

\subsection{Mirrors}
\label{sect:mirrors} Field transformations due to a mirror were
introduced in the example of Section \ref{sect:framework}. The
transformation matrix for a lossless mirror is given in
Eq.~\eqref{eq:mirror}. We now derive more complete equations for
the mirror that include losses. We ascribe a power loss $A$ to the
mirror in Fig.~\ref{fig:fig1} such that $\rho^{2}+\tau^{2}+A=1$.
The introduction of losses gives rise to an additional vacuum
field of amplitude $\sqrt{A/(1-A)}$ that is added to each input of
the mirror. The $\left(1-A\right)^{-1}$ factor accounts for part
of the loss field being lost to the mirror. This can be verified
by having shot-noise-limited fields, $\mathbf{a}$ and
$\mathbf{d}$, incident on the mirror. The field returning to the
beam block
\begin{equation}
-\rho\left(\textbf{a}+\sqrt{\frac{A}{1-A}}\textbf{v}^{(3)}\right)+
\tau\left(\textbf{d}+\sqrt{\frac{A}{1-A}}\textbf{v}^{(4)}\right)
\end{equation}
must also be at the shot noise level, such that
\begin{equation}
\rho^{2}\left(1+\frac{A}{1-A}\right)+\tau^{2}\left(1+\frac{A}{1-A}\right)
= \frac{1-A}{1-A}=1.
\end{equation}
The new equations governing the mirror are
\begin{equation}
\label{eq:mirrorloss}
\left(\begin {array}{ccc}
\textbf{b}\\
\noalign{\medskip} \textbf{c}
\end {array}\right) =
\left(\begin {array}{ccc}
-\rho&\tau\\
\noalign{\medskip} \tau&\rho
\end {array}\right)
\left(\begin {array}{ccc}
\textbf{a} + \displaystyle \sqrt{\frac{A}{1-A}}\,\textbf{v}^{(3)}\\
\noalign{\medskip} \textbf{d} + \displaystyle
\sqrt{\frac{A}{1-A}}\,\textbf{v}^{(4)}
\end {array}\right)
\end{equation}
where $\textbf{v}^{(3)}$ and $\textbf{v}^{(4)}$ are the vacuum
fluctuations that enter due to the presence of loss.

Equation~\eqref{eq:mirrorloss} may be rewritten as
\begin{equation}
\left(\begin {array}{ccc}
\textbf{b}\\
\noalign{\medskip} \textbf{c}
\end {array}\right) =
\left(\begin {array}{ccc}
-\rho&\tau\\
\noalign{\medskip} \tau&\rho
\end {array}\right)
\left(\begin {array}{ccc}
\textbf{a}\\
\noalign{\medskip} \textbf{d}
\end {array}\right)
+\sqrt{A}
\left(\begin {array}{ccc}
\textbf{v}^{(3)\prime}\\
\noalign{\medskip} \textbf{v}^{(4)\prime}
\end {array}\right),
\end{equation}
where
\begin{eqnarray}
\textbf{v}^{(3)\prime} &\equiv& \displaystyle
\sqrt{\frac{1}{1-A}}\left(-\rho\,\textbf{v}^{(3)} + \tau\,
\textbf{v}^{(4)}\right)\\
\textbf{v}^{(4)\prime} &\equiv& \displaystyle
\sqrt{\frac{1}{1-A}}\left(\tau\,\textbf{v}^{(3)} + \rho
\,\textbf{v}^{(4)}\right).
\end{eqnarray}
$\textbf{v}^{(3)\prime}$ and $\textbf{v}^{(4)\prime}$ are
uncorrelated vacuum fields in this representation. We can subsequently
write the mirror's contribution to Eq.~\eqref{general_equation} as
\begin{equation}
\label{mirror:eq:format}
\left(
\begin{array}{cccc}
-\rho & -1 & 0 & \tau \\
\tau & 0 & -1 & \rho
\end{array}
\right)
\left(\begin {array}{cccc}
\textbf{a}\\
\noalign{\medskip}
\textbf{b}\\
\noalign{\medskip}
\textbf{c}\\
\noalign{\medskip}
\textbf{d}
\end {array}\right)
=
\left(
\begin{array}{cc}
-\sqrt{A}\mathbf{v}^{(3)\prime} \\
-\sqrt{A}\mathbf{v}^{(4)\prime}
\end{array}
\right)\,.
\end{equation}
This method may also be used to inject losses in beamsplitters or
cavities.

\subsection{Free space propagation}
\begin{figure}[t]
\includegraphics[width=0.45\textwidth]{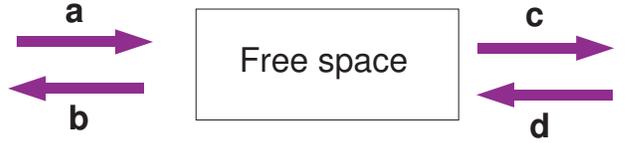}
\caption{\label{fig:cavity} The fields entering and exiting a
region of free space are shown. Propagation operators are
characterized by the propagation distance (and orientation
relative to the source polarization, in the case of the GW
signal).}
\end{figure}

Since optical cavities are present in virtually all optical
configurations of gravitational-wave interferometers, we must give
a transformation matrix for them as an element of our arbitrary
optical train. To do so we introduce an operator to transform the
field as it propagates through free space between any two other
optical elements (in the case of an optical cavity, these would be
mirrors). Using the convention of Fig.~\ref{fig:cavity}, the
matrix for propagation through a length $L$ transforms input
fields $\textbf{a}$ and $\textbf{d}$ according to
\begin{equation}
\left(\begin {array}{cccc}
\textbf{b}\\
\noalign{\medskip}
\textbf{c}
\end {array}\right) =
\textbf{M}_{\rm Prop}\, \left(\begin {array}{cccc}
\textbf{a}\\
\noalign{\medskip}
\textbf{d}
\end {array}\right)
\dfrac{}{}\end{equation} where the matrix for the propagator is
\begin{equation}
\textbf{M}_{\rm Prop} \equiv e^{i\phi}\left(\begin {array}{cccc}
0 & \mathbf{R}_{\Theta}\\
\noalign{\medskip}
\mathbf{R}_{\Theta} & 0
\end {array}\right).
\end{equation}

\noindent Here
\begin{eqnarray}
\label{eq:Theta:cavity}
%\Theta &\equiv& \left[\frac{\omega L}{c}\right]_{\mbox{\scriptsize
%mod
%    $2\pi$}}\,, \\
\Theta &\equiv& \frac{\omega L}{c}\,, \\
\label{eq:phi}
\phi &\equiv& \frac{\Omega L}{c}\,,
\end{eqnarray}
are the one-way phase shift on the carrier light at frequency,
$\omega$, and on modulation sidebands at frequency, $\Omega$,
respectively, and
\begin{equation}
\mathbf{R}_{\Theta} \equiv \left(\begin {array}{cccc}
\cos\Theta&-\sin\Theta\\
\noalign{\medskip}
\sin\Theta&\cos\Theta
\end {array}\right)
\end{equation}
is the rotation operator on quadrature fields.

\subsection{Beamsplitters}
\begin{figure}[t]
\includegraphics[height=7cm]{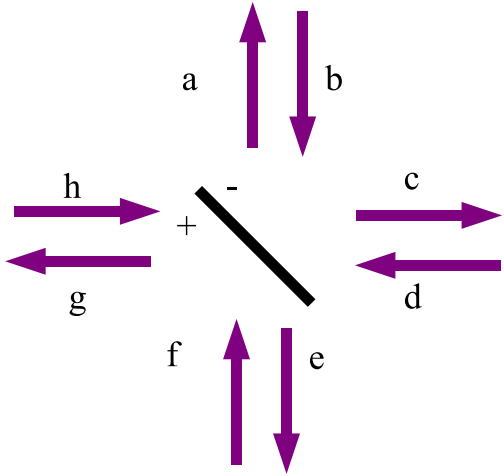}
\caption{\label{fig:bs} Treating the beamsplitter as a four-port
device, definitions for the fields, including sign conventions,
are shown.}
\end{figure}

Another essential optical element of an interferometer is the
beamsplitter. We consider a beamsplitter with amplitude
reflectivity and transmissivity, $\rho$ and $\tau$, respectively.
The beamsplitter transforms the input fields, shown in Figure
\ref{fig:bs}, according to the matrix equation

\begin{equation}
\label{eq:bs:norp}
\left(\begin {array}{cccc}
\textbf{a}\\
\noalign{\medskip}
\textbf{c}\\
\noalign{\medskip}
\textbf{e}\\
\noalign{\medskip}
\textbf{g}
\end {array}\right) =
\textbf{M}_{BS}
\left(\begin {array}{cccc}
\textbf{d}\\
\noalign{\medskip}
\textbf{b}\\
\noalign{\medskip}
\textbf{h}\\
\noalign{\medskip}
\textbf{f}
\end {array}\right)
\end{equation}
where
\begin{equation}
\textbf{M}_{BS} \equiv
\left(\begin {array}{cccc}
-\rho&0&0&\tau\\
\noalign{\medskip}
0&-\rho&\tau&0\\
\noalign{\medskip}
0&\tau&\rho&0\\
\noalign{\medskip}
\tau&0&0&\rho
\end {array}\right).
\end{equation}

In presence of optical loss, assuming $\rho^2+\tau^2+A=1$, and
going through similar arguments to Sec.~\ref{sect:mirrors}, we
simply add a column vector of vacuum fields
$-\sqrt{A}\mathbf{v}^{(i)}$ ($i=1,2,3,4$) onto the right-hand side
of Eq.~\eqref{eq:bs:norp}.

\subsection{Correlators}
The correlator module of the simulation allows for the inclusion
of squeezed light or vacuum fields in the interferometer. It is
essentially a one-way device: only fields entering from one
direction are transformed; fields entering from the other
direction pass through the correlator unmodified. Taking
$\textbf{a}$ to be the input field, the field at the output of the
correlator, $\textbf{b}$, is defined by
\begin{equation}
\textbf{b} = \textbf{S}\left(r,\phi\right) \textbf{a},
\end{equation}
where $\textbf{S}\left(r,\phi\right)$ is the squeeze operator with
squeeze factor $r$ and squeeze angle $\phi$:
\begin{eqnarray}
\textbf{S}\left(r,\phi\right) \equiv \nonumber \hspace{6cm}\\
\left(\begin {array}{cccc}
\cosh r + \sinh r \cos 2 \phi&\sinh r \sin 2 \phi\\
\noalign{\medskip}
\sinh r \sin 2 \phi&\cosh r - \sinh r \cos 2 \phi
\end {array}\right).
\end{eqnarray}

\section{Radiation Pressure}
\label{sect:radpres}

Radiation pressure plays an important role in interferometers
operating close to or beyond the SQL, since quantum back-action
noise must be taken into account. Moreover, radiation-pressure
effects can also modify the dynamics of these
interferometers~\cite{BC2}. Sideband quadrature fields create
amplitude modulations to the carrier field, and the associated
power modulation drives the motion of optical elements, which, in
turn, phase modulates the carrier, thereby creating sideband
quadrature fields.

Details of this sideband-to-sideband conversion depend on the
phases (this determines which quadrature gets converted into
which) and amplitudes (this determines the conversion strength) of
the carrier field propagating in different parts of the
interferometer. Therefore, it is necessary to separate the fields
into carrier ($\Omega=0$) and sideband ($\Omega\ne 0$) components
at this point. The radiation pressure force due to the carrier
field itself is a time independent force and can be ignored (in
reality they will be balanced by a static force exerted on the
optical elements, e.g., the pendulum restoring force on a
suspended mirror). The effect of interest is the time-dependent
part of the force, due to sideband components, which will be the
subject of this section. As a foundation, we must first of all
calculate the phase and amplitude of the carrier fields at each
location. But this we can already do by building the general
equation~\eqref{general_equation} out of input-output relations of
static optical elements, which have already been derived in
Sec.~\ref{sect:matrices}, and solving it.

Before incorporating radiation pressure into the treatment of
specific systems, let us study the electromagnetic momentum flux
carried by optical fields in the two-photon formalism. In
quadrature representation, we decompose the total quadrature field
$\mathbf{E}^{\rm total}_j$ (here $\textbf{E}_j$ can be
$\textbf{a}$, $\textbf{b}$, $\textbf{c}$ or $\textbf{d}$ for the
configuration in Fig.~\ref{fig:fig1}) into the following two
terms:

\beq \label{decomp} \mathbf{E}^{\rm total}_j =\mathbf{E}^{\rm
carrier}_j+\mathbf{E}^{\rm sb}_j\,. \eeq

The monochromatic carrier field in Eq.~\eqref{decomp} can be
written more explicitly in terms of power $I_{j}$, phase
$\theta_{j}$ and effective beam area $\cal A$ as \beq
\mathbf{E}_j^{\rm carrier} = \sqrt{\frac{8\pi I_j}{\mathcal{A}c}}
\left( \cvec{\cos\theta_j}{\sin\theta_j}\right), \eeq while the
sideband field can be written as an integral over all sideband
frequencies:

\beq \mathbf{E}^{\rm sb}_j(t) =\sqrt{\frac{4\pi\hbar\omega}{{\cal
A}c}} \int _0^{+\infty}
\frac{d\Omega}{2\pi}\Big[\mathbf{j}(\Omega)e^{-i\Omega t}
+H.c.\Big]. \eeq The total momentum flow carried by the field is
\beq \frac{\mathcal{A}}{4\pi}\left(\mathbf{E}_{j}^{\rm
carrier}+\mathbf{E}_{j}^{\rm sb}\right)^{2}. \eeq Removing the
static (dc) and optical frequency ($\omega$) components, the
Fourier transform of the time-averaged (over a time scale much
shorter than the GW period, but much longer than $1/\omega$) ac
momentum flow carried by this field is
\beq
\label{eq:momflux}
\dot{P_{j}}(\Omega)=
\sqrt{\frac{\hbar\omega}{c^2}}\mathbf{D}_j^T \mathbf{j}(\Omega)
\,, \eeq where we have defined \beq \mathbf{D}_j \equiv
\sqrt{\frac{\mathcal{A}c}{4\pi}\,}\mathbf{E}_j^{\rm carrier} =
\sqrt{2 I_j}\left(\cvec{\cos\theta_j}{\sin\theta_j}\right)\,
\eeq
as the carrier quadrature field, and $\textbf{j}(\Omega)$ is the
sideband component at angular frequency $\Omega$.

In the remainder of this Section we derive explicit input-output
relations for mirrors and beamsplitters, including radiation
pressure effects. Our results will be more general than previously
obtained results by allowing the carrier fields incident from
different ports to have different phases.

\subsection{Mirrors}

Let us once again consider the mirror in Fig.~\ref{fig:fig1}.
Assuming that the mirror behaves as a free particle with mass $M$
when no radiation-pressure forces are exerted (valid for suspended
mirrors when frequencies greater than the pendulum resonant
frequency are considered), the Fourier transform for the equation
of motion for the mirror is
\begin{equation}
-M\Omega^{2} X = \sum_{j} \eta_{j}\dot{P_{j}}
\end{equation}
where X is the displacement of the mirror {\it induced by all the
sideband fields} ($X$ is positive to the left in
Fig.~\ref{fig:fig1}, and the $j$ refer to $\mathbf{a,b,c,d}$). The
summation is performed over all the fields entering and exiting
the mirror; the coefficients $\eta_{a}=\eta_{b}=-1$ and
$\eta_{c}=\eta_{d}=1$ account for the directions of propagation.
The displacement of the mirror due to the radiation pressure
forces, $X$, can be written explicitly as [see
Eq.~\eqref{eq:momflux}] \bea \label{eq:mirror_ba} X =
\frac{1}{M\Omega^2}\sqrt{\frac{\hbar\omega}{c^2}} &\bigg[&
\left(\rvec{\mathbf{D}_a^T}{-\mathbf{D}_d^T}\right)
\left(\cvec{\mathbf{a}}{\mathbf{d}}\right)\nonumber \\
&+& \left(\rvec{\mathbf{D}_b^T}{-\mathbf{D}_c^T}\right)
\left(\cvec{\mathbf{b}}{\mathbf{c}}\right) \bigg]\,. \eea
\noindent Given a (time-dependent) displacement $X(t)$ of the
mirror, the input-output relation can be written as (if $\dot{X}
\ll c$)
\begin{subequations}
\bea
\label{eq:mirror_sig1}
E^{\rm total}_b(t)& =& -\rho E_a^{\rm total}\left[t + \frac{2X(t)}{c}\right]
+\tau E^{\rm total}_d (t)\, \quad \\
\label{eq:mirror_sig2}
E^{\rm total}_c(t)& =&  \tau E_a^{\rm total} (t) + \rho E_d^{\rm total}
\left[t-\frac{2X(t)}{c}\right]\,.
\eea
\end{subequations}
$c$ in the argument of $E^{\rm total}_j$ for the $j$-th field is
the speed of light and should be distinguished from $c$ in the
subscript of $E^{\rm total}_j$, which refers to the field
$\textbf{c}$. In quadrature representation, to leading order in
$X$ and in the sideband field amplitudes, we have,
\bea
\label{eq:signal}
&&E^{\rm total}_j \left[t \mp \frac{2X(t)}{c}\right] \nonumber  \\
&\Leftrightarrow &\mathbf{E}_j^{\rm carrier} + \mathbf{E}_j^{\rm
sb}(t) \pm \frac{2\,\omega \, X(t)}{c}\,\mathbf{R}_{\Theta =
\pi/2}\,\mathbf{E}_j^{\rm carrier}\,,
\qquad\nonumber \\
&=&\mathbf{E}_j^{\rm carrier} + \mathbf{E}_j^{\rm sb}(t) \mp
\frac{2\,\omega \,X(t)}{c}\left[\mathbf{E}_j^{\rm
carrier}\right]^*\,.
\eea

\begin{figure}[t]
\includegraphics[width=0.3\textwidth]{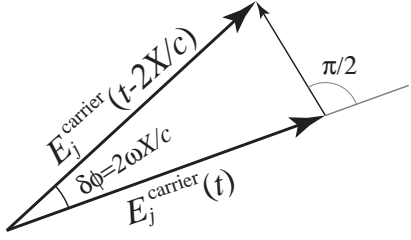}
\caption{\label{fig:signal} Here we show that the phase modulation
sideband generated by the radiation pressure force is
perpendicular to the carrier field, which is why the generated
signal has a $\textbf{D}^{*}$ dependence.}
\end{figure}

Here $^*$ refers to a rotation by
$\pi/2$, as described by $-\mathbf{R}_{\Theta=\pi/2}$ in
Eq.~\eqref{eq:signal}. Accordingly, for any quadrature field
$\mathbf{v}$, we define \beq \mathbf{v}^* \equiv
\left(\cvec{v_2}{-v_1}\right)\,,\quad\mbox{for}\;\mathbf{v}=
\left(\cvec{v_1}{v_2}\right)\,. \eeq
Equation~\eqref{eq:signal} implies that time delays, or phase modulations, create sideband quadratures orthogonal to the carrier, as illustrated in terms of phasors in Fig.~\ref{fig:signal}.
The sideband part, i.e., the ac components in
Eqs.~\eqref{eq:mirror_sig1} and~\eqref{eq:mirror_sig2}, can be
obtained using Eq.~\eqref{eq:signal}:
\beq \label{eq:mirror_out}
\left(\cvec{\mathbf{b}}{\mathbf{c}}\right) =\mathbf{M}_{\rm
mirror} \left(\cvec{\mathbf{a}}{\mathbf{d}}\right)
-\frac{2\,\rho\,\omega \, X}{c\,\sqrt{\hbar\,\omega}}
\left(\cvec{\mathbf{D}_a^*}{\mathbf{D}_d^*}\right)\,. \eeq

Inserting Eq.~\eqref{eq:mirror_ba} into Eq.~\eqref{eq:mirror_out}
gives
\bea \label{eq:RPM} &&\left[\mathbf{I}+\Pi
\left(\cvec{\mathbf{D}_a^*}{\mathbf{D}_d^*}\right)
\left(\rvec{\mathbf{D}_b^T}{-\mathbf{D}_c^T}\right)\right]
\left[\cvec{\mathbf{b}}{\mathbf{c}}\right] \nonumber \\
&=& \left[\mathbf{M}_{\rm mirror}-\Pi
\left(\cvec{\mathbf{D}_a^*}{\mathbf{D}_d^*}\right)
\left(\rvec{\mathbf{D}_a^T}{-\mathbf{D}_d^T}\right)\right]
\left[\cvec{\mathbf{a}}{\mathbf{d}}\right]\,.\qquad\,. \eea where
\beq \Pi\equiv \frac{2\,\rho\,\omega}{M\,\Omega^2\,c^2} \eeq is a
quantity with units of inverse power or ${\rm W}^{-1}$. [For lossy
mirrors with $\rho^2+\tau^2+A=1$, we simply insert a column vector
$-\sqrt{A}\mathbf{v}^{(i)}$, $i=1,2$ onto the right-hand sides of
Eq.~\eqref{eq:mirror_out} and \eqref{eq:RPM},
Cf.~Sec.~\ref{sect:mirrors}.]

To solve for $\textbf{b}$ and $\textbf{c}$, the matrix on the left
hand side of Eq.~\eqref{eq:RPM} must be inverted. It is
straightforward to find a complete set of eigenvectors for this
matrix, they are: \beq \left[\mbox{\boldmath$\xi$}_1
,\mbox{\boldmath$\xi$}_2 ,\mbox{\boldmath$\xi$}_3
,\mbox{\boldmath$\xi$}_4\right]= \left[
\left(\cvec{\mathbf{D}_b^*}{0}\right),
\left(\cvec{0}{\mathbf{D}_c^*}\right),
\left(\cvec{\mathbf{D}_c}{\mathbf{D}_b}\right),
\left(\cvec{\mathbf{D}_a^*}{\mathbf{D}_d^*}\right)\right]\,. \eeq
Since the first three vectors are orthogonal to
$\left(\rvec{\mathbf{D}_b^T}{\hspace{0.1cm}-\mathbf{D}_c^T}\right)$,
the three corresponding eigenvalues are
$\lambda_1=\lambda_2=\lambda_3=1$; the last eigenvalue is \bea
\lambda_4&=&1+\Pi \left[\mathbf{D}_b^T
\mathbf{D}_a^*-\mathbf{D}_c^T
\mathbf{D}_d^*\right] \nonumber \\
&=&1 + 2 \, \tau \, \Pi \,\mathbf{D}_d^T \mathbf{D}_a^* \nonumber \\
&=& 1 + \frac{8\,\rho\,\tau\,\omega\, \sqrt{I_a\,
I_d}}{M\,\Omega^2\,c^2}\sin(\theta_a - \theta_d)\,. \eea Inverting
the eigenvalue $\lambda_4$ yields a pair of resonant frequencies
at \beq \pm {\Omega}_{\rm M}=
\pm\left[\frac{-8\,\rho\,\tau\,\omega\,\sqrt{I_a\,
I_d}}{M\,c^2}\sin(\theta_a - \theta_d)\right]^{1/2}\,. \eeq
Physically, this resonance comes about because the sideband fields
generated by mirror motion can exert radiation pressure back onto
the mirror. Let us for a moment consider classical motion of the
mirror. As was mentioned after Eq.~\eqref{eq:mirror_out}, for any
given input carrier field, the sideband field generated upon
reflection from the moving mirror is $\pi/2$ phase shifted
relative to the input carrier, so the sideband will not beat with
the reflected carrier to induce any force on the mirror [see
Eq.~\eqref{eq:momflux}] --- force can only be induced by beating
this motion-induced sideband field with the transmitted carrier,
which must have non-zero amplitude and must have a phase
difference other than $\pi/2$ relative to the sideband. This
explains why the resonant frequency vanishes if either $\rho=0$ or
$\tau = 0$ (no reflected or transmitted field), or if
$\theta_a-\theta_d=N\pi$ (no phase difference between the two
input fields).

When the two input carrier fields, $\mathbf{D}_a$ and
$\mathbf{D}_d$, have the same phase (or differ by $N\pi$), the
phasors corresponding to $\mathbf{D}_a$, $\mathbf{D}_b$,
$\mathbf{D}_c$ and $\mathbf{D}_d$ all become parallel to each
other. This is true for almost all interferometers that have been
treated explicitly analytically. This case is rather special from
a mathematical point of view, since the matrix we are inverting
does {\it not} have a complete set of eigenvectors. Fortunately,
the inverse is just \bea &&\left[\mathbf{I} + \Pi\,
\left(\cvec{\mathbf{D}_a^*}{\mathbf{D}_d^*}\right)
\left(\rvec{\mathbf{D}_b^T}{-\mathbf{D}_c^T}\right)\right]^{-1}\nonumber \\
&=& \left[\mathbf{I} - \Pi\,
\left(\cvec{\mathbf{D}_a^*}{\mathbf{D}_d^*}\right)
\left(\rvec{\mathbf{D}_b^T}{-\mathbf{D}_c^T}\right)\right]\,,\quad
{\mbox{if}}\; \mathbf{D}_a \parallel \mathbf{D}_d\,;\qquad \eea
since
\beq
\left[\left(\cvec{\mathbf{D}_a^*}{\mathbf{D}_d^*}\right)
\left(\rvec{\mathbf{D}_b^T}{-\mathbf{D}_c^T}\right)\right]^2=0\,,\quad
{\mbox{if}}\; \mathbf{D}_a \parallel \mathbf{D}_d\,.
\eeq
(This identity originates from the fact that the
sideband field is orthogonal to the carrier field about which it
is generated.) Using this fact, we can further simplify the input-output relation
to
\begin{widetext}
\bea
\left(\cvec{\mathbf{b}}{\mathbf{c}}\right)
=
\Bigg[\mathbf{M}_{\rm mirror}
-
2\rho\Pi
\left(\cvec{\mathbf{D}_a^*}{\mathbf{D}_d^*}\right)
\left(\rvec{\mathbf{D}_a^T}{-\mathbf{D}_d^T}\right)
\left(\begin{array}{cc}
\rho & -\tau \\
\tau & \rho
\end{array}
\right)
\Bigg]
\left(\cvec{\mathbf{a}}{\mathbf{d}}\right)\,,\quad {\mbox{if}}\;
\mathbf{D}_a \parallel \mathbf{D}_d\,.\qquad
\eea

[Here for simplicity we have assumed the mirror to be lossless.]
In practice, although Eq.~\eqref{eq:RPM} does not give the output
fields $\mathbf{b}$ and $\mathbf{c}$ explicitly in terms of the
input fields $\mathbf{a}$ and $\mathbf{d}$, it can be incorporated
to the matrix $\mathcal{M}$ (and into $\mathbf{u}^{(i)}$, in
presence of optical losses) without any trouble
[cf.~Eq.~\eqref{general_equation}]: its inversion will take place
{\it automatically} when $\mathcal{M}^{-1}$ is calculated.
[However, doing so will make it impossible to have $-1$ all along
the diagonal of $\mathcal{M}$.] Alternatively, the variable $X$
may be added to our system of variables,  with
Eq.~\eqref{eq:mirror_ba} providing the additional equation
necessary. The equations governing a mirror may then  be replaced
with Eq.~\eqref{eq:mirror_out} to include the dependence on $X$.
In this way, the $-1$ diagonal components are preserved, without
the need to invert additional matrices.

\subsection{Beamsplitter}
\label{sect:bsradpress}

Referring to the fields shown in Fig.~\ref{fig:bs}, the
displacement due to radiation pressure forces on a beamsplitter
(normal to its reflective face) is \bea \label{eq:motion_BS}
X_{\rm N}= \frac{X_x+X_y}{\sqrt{2}}
 =\frac{1}{M\Omega^2}\sqrt{\frac{\hbar\omega}{2c^2}}
\left[
\left(\begin{array}{cccc}
\mathbf{D}_a^T & \mathbf{D}_c^T  & -\mathbf{D}_e^T  & -\mathbf{D}_g^T
\end{array}\right)
\left(\begin{array}{c}
\mathbf{a} \\
\mathbf{c} \\
\mathbf{e} \\
\mathbf{g}
\end{array}\right)
+\left(\begin{array}{cccc}
\mathbf{D}_d^T & \mathbf{D}_b^T  & -\mathbf{D}_h^T  & -\mathbf{D}_f^T
\end{array}\right)
\left(\begin{array}{c}
\mathbf{d} \\
\mathbf{b} \\
\mathbf{h} \\
\mathbf{f}
\end{array}\right)\right]\,.\qquad
\eea
where $X_{x}$ is the displacement along the x-axis and
$X_{y}$ is the displacement along the y-axis. Similar to the case
of a cavity mirror, this motion induces phase fluctuations on the
impinging fields upon reflection, and introduces additional terms
in the input-output relation. Following a procedure similar to the
one with which we obtain Eq.~\eqref{eq:mirror_out}, we get \bea
\label{eq:signal_BS} \left(
\begin{array}{c}
\mathbf{a} \\
\mathbf{c} \\
\mathbf{e} \\
\mathbf{g}
\end{array}
\right)
=\mathbf{M}_{\rm BS}
\left(
\begin{array}{c}
\mathbf{d} \\
\mathbf{b} \\
\mathbf{h} \\
\mathbf{f}
\end{array}\right)
-\frac{\sqrt{2}\rho\omega X_{\rm N}}{c\sqrt{\hbar\omega}}
\left(
\begin{array}{c}
\mathbf{D}_d^* \\
\mathbf{D}_b^* \\
\mathbf{D}_h^* \\
\mathbf{D}_f^*
\end{array}\right)\,.
\eea
Inserting Eq.~\eqref{eq:motion_BS} into Eq.~\eqref{eq:signal_BS} gives
\bea
\label{eq:RPBS}
\left[\mathbf{I}+\frac{\Pi}{2}
\left(
\begin{array}{c}
\mathbf{D}_d^* \\
\mathbf{D}_b^* \\
\mathbf{D}_h^* \\
\mathbf{D}_f^*
\end{array}\right)
\left(\begin{array}{cccc}
\mathbf{D}_a^T & \mathbf{D}_c^T  & -\mathbf{D}_e^T  & -\mathbf{D}_g^T
\end{array}\right)
\right]
\left(
\begin{array}{c}
\mathbf{a} \\
\mathbf{c} \\
\mathbf{e} \\
\mathbf{g}
\end{array}
\right)
=
\left[
\mathbf{M}_{\rm BS}
-\frac{\Pi}{2}
\left(\begin{array}{c}
\mathbf{D}_d^* \\
\mathbf{D}_b^* \\
\mathbf{D}_h^* \\
\mathbf{D}_f^*
\end{array}\right)
\left(\begin{array}{cccc}
\mathbf{D}_d^T & \mathbf{D}_b^T  & -\mathbf{D}_h^T  & -\mathbf{D}_f^T
\end{array}\right)\right]
\left(
\begin{array}{c}
\mathbf{d} \\
\mathbf{b} \\
\mathbf{h} \\
\mathbf{f}
\end{array}
\right) \eea Equation~\eqref{eq:RPBS} is quite similar in nature
to Eq.~\eqref{eq:RPM}; optical losses can also be incorporated in
a similar fashion, by adding $-\sqrt{A}\mathbf{v}^{(i)}$,
$i=1,2,3,4$ on to its right-hand side, where $\rho^2+\tau^2+A=1$.
Again, in the generic case where \beq \left(\begin{array}{cccc}
\mathbf{D}_a^T & \mathbf{D}_c^T  & -\mathbf{D}_e^T  &
-\mathbf{D}_g^T
\end{array}\right)
\left(\begin{array}{c}
\mathbf{D}_d^* \\
\mathbf{D}_b^* \\
\mathbf{D}_h^* \\
\mathbf{D}_f^*
\end{array}\right)
\neq 0\,, \eeq the matrix on the LHS of Eq.~\eqref{eq:RPBS} has
eight linearly independent eigenvectors, of which seven have unit
eigenvalue, while the eighth has \bea \lambda_8 &=&
1+\frac{\Pi}{2} \left(\begin{array}{cccc} \mathbf{D}_a^T &
\mathbf{D}_c^T  & -\mathbf{D}_e^T  & -\mathbf{D}_g^T
\end{array}\right)
\left(\begin{array}{c}
\mathbf{D}_d^* \\
\mathbf{D}_b^* \\
\mathbf{D}_h^* \\
\mathbf{D}_f^*
\end{array}\right) \nonumber \\
&=&
1+\tau \Pi (\mathbf{D}_f^T \mathbf{D}_d^*+ \mathbf{D}_h^T \mathbf{D}_b^*)\nonumber \\
&=&1+\frac{4\rho\tau\omega_0}{M\Omega^2c^2}\left[\sqrt{I_f
I_d}\sin(\theta_f-\theta_d) +\sqrt{I_h
I_b}\sin(\theta_h-\theta_b)\right] \,, \eea which corresponds to
an opto-mechanical resonance at angular frequency \bea
\pm\Omega_{\rm BS} &=&\pm \bigg\{-\frac{4\rho\tau\omega}{Mc^2}
\Big[\sqrt{I_h I_b}\sin(\theta_h-\theta_b)+\sqrt{I_f
I_d}\sin(\theta_f-\theta_d)\Big] \bigg\}^{1/2}\,.\quad \eea In the
special case of \beq \left(\begin{array}{cccc} \mathbf{D}_a^T &
\mathbf{D}_c^T  & -\mathbf{D}_e^T  & -\mathbf{D}_g^T
\end{array}\right)
\left(\begin{array}{c}
\mathbf{D}_d^* \\
\mathbf{D}_b^* \\
\mathbf{D}_h^* \\
\mathbf{D}_f^*
\end{array}\right)
= 0\,, \eeq i.e., all input carrier fields are in phase with each
other (modulo $\pi$) we get \bea \left(
\begin{array}{c}
\mathbf{a} \\
\mathbf{c} \\
\mathbf{e} \\
\mathbf{g}
\end{array}
\right)
=\left[\mathbf{M}_{\rm BS} -\rho\Pi
\left(\begin{array}{c}
\mathbf{D}_d^* \\
\mathbf{D}_b^* \\
\mathbf{D}_h^* \\
\mathbf{D}_f^*
\end{array}\right)
\left(\begin{array}{cccc}
\mathbf{D}_d^T & \mathbf{D}_b^T  & -\mathbf{D}_h^T  & -\mathbf{D}_f^T
\end{array}\right)
\left(
\begin{array}{cccc}
\rho & & & -\tau\\
 & \rho & -\tau & \\
 & \tau & \rho & \\
 \tau & & & \rho
 \end{array}
\right)
\right]\left(
\begin{array}{c}
\mathbf{d} \\
\mathbf{b} \\
\mathbf{h} \\
\mathbf{f}
\end{array}
\right)\,,\quad {\mbox{if}}\; \mathbf{D}_b \parallel
\mathbf{D}_d\parallel \mathbf{D}_f\parallel \mathbf{D}_h\,.\qquad.
\eea
\end{widetext}

For simplicity, we assume the beamsplitter to be lossless in the
above equation. This is particularly true for the {\it
beamsplitter} in Michelson- and Sagnac-type GW
interferometers~\cite{BS}. Similar to the case of the mirror, for
the purposes of simulation, we incorporate the position of the
beamsplitter as an additional variable in $\mathcal{M}$, in order
to preserve the $-1$ diagonal elements and to avoid the inversion
of additional matrices.

\section{Gravitational wave signal and the output field}
\label{sect:gwsignal}

\subsection{GW contribution}
In our set of optical elements, only optical cavities have
significant propagation distances, so we model the effect of GWs
by introducing a phase shift to the carrier light as it passes
through a cavity. To calculate the propagation of these fields,
all that must be done is to add a source term in the equation
governing the cavity. Refering to the fields in
Fig.~\ref{fig:cavity}, the cavity field becomes

\bea
\label{cavity_gw}
\mathbf{c}&=&e^{i\phi}\mathbf{R}_{\Theta}\mathbf{a}
-\eta \frac{\omega L h}{2 c\sqrt{\hbar\omega}} \mathbf{D}_c^* \nonumber \\
&=& \mathbf{R}_{\Theta}\left[e^{i\phi}\mathbf{a}-\eta \frac{\omega
\,L \,h}{2 \,c\,\sqrt{\hbar\,\omega}} \mathbf{D}_a^*\right]\, \eea
where $h$ is the Fourier transform of the GW amplitude. An
$h$-dependent term is also added to the equation relating
$\textbf{b}$ and $\textbf{d}$ using $\mathbf{D}_{d}^*$ in place of
$\mathbf{D}_{a}^*$. The parameter $\eta$ takes values from $-1$ to
$1$ depending on the orientation of the cavity and the
polarization state of the incoming GW. For example, for a linearly
polarized incoming GW, and for an optimally aligned Michelson
interferometer, we have $\eta=1$ for one and $ -1$ for the other.

It is straightforward to incorporate Eq.~\eqref{cavity_gw} into
the general equation Eq.~\eqref{general_equation}. In particular,
the term containing $h$ on RHS contributes to the GW part of the
general input field $\mathbf{u}$, i.e., to the third term of
Eq.~\eqref{general_u}, with
\begin{equation}
\mathbf{H} = -\eta \frac{\omega L }{2 c\sqrt{\hbar\omega}}\,\mathbf{D}_c^*,
\end{equation}

\subsection{Photodetection: signal and noise}

For our purposes, the photodetector serves two roles: first, it
represents an open port, from which vacuum fluctuations enter the
interferometer; second, it determines the measurement point. For
the former, the input-output relation of a photodetector, as it
contributes to the matrix $\mathcal{M}$ and the generalized input
vector $\mathbf{u}^{(i)}$,  is trivial and has been discussed in
Sec.~\ref{subsect:blockpd}. Here we focus on the latter. At zero
frequency, there is only contribution to $\mathbf{b}$ from the
carrier laser, while at non-zero sideband frequencies, the
detected fields at a photodetector comprise three components: the
gravitational-wave signal, classical laser noise, and noise due to
vacuum fluctuations in the detected mode. The outgoing field being
detected, $\textbf{b}$, has the general form [see
Eqs.~\eqref{general_equation} and \eqref{general_u}]:
\begin{eqnarray}
\mathbf{b} &=& \sum_{i}\left[\mathcal{M}^{-1}\right]_{bi}\left[
\mathbf{v}^{(i)} + \mathbf{l}^{(i)} + \mathbf{H}^{(i)}h
\right] \nonumber \\
&\equiv& \sum_{i} \mathcal{T}_{bi}\left[
\mathbf{v}^{(i)} + \mathbf{l}^{(i)} + \mathbf{H}^{(i)}h
\right] \,.
\end{eqnarray}
The summation is performed over all fields. We note that
contributions to  $\mathbf{v}^{(i)}$ exist only for fields that
emerge from beam blocks or lossy optical elements, those to
$\mathbf{l}^{(i)}$ exist only for the field that emerges from the
laser, and those to $\mathbf{H}^{(i)}$ only for fields that emerge
from cavities.

We suppose homodyne detection at quadrature angle, $\zeta$, is performed such
that the measured field is
\begin{equation}
\textbf{b}_{\zeta} = b_{1} \cos \zeta + b_{2} \sin \zeta.
\end{equation}
For a complete simulation, $\zeta$ should be the phase of the
carrier that emerges at this port. However, in theoretical
studies, we could also assign another value to $\zeta$, assuming
that the local-oscillator phase is modified by some other means
the simulation does not address.

For the detected field, the quantum noise spectral density  is
(see, e.g., Sec.~III of Ref.~\cite{KLMTV})
\begin{eqnarray}
\label{eq:qnoise}
\left(N^{2}_{Q}\right)_b
=
\sum_i \left[\begin{array}{cc} \cos\zeta & \sin\zeta \end{array}\right]
\mathcal{T}_{bi}   S_{\mathbf{v}_{i}} \mathcal{T}^\dagger_{bi}
\left[\begin{array}{c} \cos\zeta \\ \sin\zeta \end{array}\right]\,.
\end{eqnarray}
Because $\mathbf{v}_i$ is always proportional to a vacuum field,
we have used $S_{\mathbf{v}_i}$ to denote the noise spectral
density which is identical for all its quadratures. Here we have
added the {\it power} of different loss contributions, since we
assume the vacuum fields to be independent to each other. In
general, laser noise is neither quantum-limited, nor are the
magnitudes of phase and amplitude fluctuations equal; there could
also be correlations between the laser amplitude and phase noise,
even as the laser field enters the system. Taking these into
account, we have a laser noise spectral density of
\begin{equation}
\left(N^2_L\right)_b
=
\left[\begin{array}{cc} \cos\zeta & \sin\zeta \end{array}\right]
\mathcal{T}_{bl}   \mathcal{S}_{L} \mathcal{T}^\dagger_{bl}
\left[\begin{array}{c} \cos\zeta \\ \sin\zeta \end{array}\right]\,,
\end{equation}
where $l$ corresponds to the input laser field, and
\begin{equation}
\mathcal{S}_L
\equiv
\left[
\begin{array}{cc}
S_{11} & S_{12}\\
S_{12} & S_{22}
\end{array}
\right]
\end{equation}
describes noise of the laser as it first enters the
interferometer, with $S_{11(22)}$ being the noise spectral density
of the first (second) quadrature,  and $S_{12}$ the cross spectral
density between the two quadratures. [In the usual convention of
having input laser in the first quadrature, $1$ corresponds to the
amplitude quadrature, hence amplitude noise, while $2$ corresponds
to the phase quadrature, hence phase noise.] The transfer function
for the GW signal is
\begin{equation}
H_b \equiv
\sum_{i}
\left[
\begin{array}{cc}
\cos\zeta & \sin\zeta
\end{array}
\right]
\mathcal{T}_{bi} \mathbf{H}^{(i)}\,.
\end{equation}
Note that GW contributions from different parts of the system add up coherently.
The displacement (strain) noise spectral density from quantum noise is then given
by
\begin{equation}
S_{h} =
\frac{N_Q^2+N_L^2}{|H|^2}\,.
\end{equation}

\section{Application to a complex interferometer}
\label{sect:ifoexample}

The mathematical formulation described in Sections
\ref{sect:framework} through \ref{sect:gwsignal} was encoded into
a simulation program written in C++. In this section we describe
tests of the simulation code for a complex interferometer
configuration, where the simulation results were compared with
analytic calculations.

\begin{figure}[b]
\includegraphics[height=3.9cm]{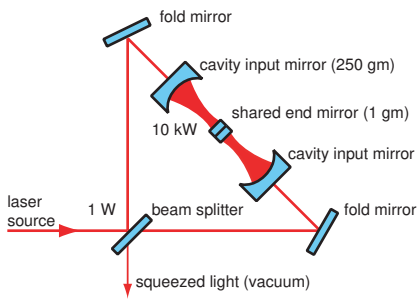}
\caption[Figure]
{\label{fig:ponderomotive_layout} Schematic of a an interferometer
designed to extract ponderomotively squeezed light due to
radiation-pressure-induced motion of the ultra-light shared
mirror. Light from a highly intensity- and frequency-stabilized
laser source is incident on the beamsplitter. High-finesse
Fabry-Perot cavities in the arms of the Michelson interferometer
are used to build up the carrier field incident on the end mirrors
of the cavity, which are a single mechanical object. }
\end{figure}

\begin{figure}[ht]
\centerline{\includegraphics[width=0.45\textwidth]{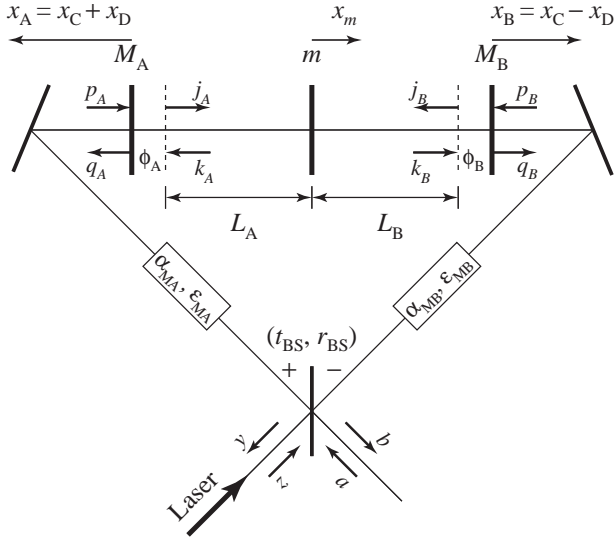}}
\caption{Optical fields propagating in the interferometer, and
modes of motion of the mirrors. In particular,
$\tinysub{\alpha}{MA,B}$ and $\tinysub{\epsilon}{MA,B}$ are
artificial detunings and losses one can add to the two arms of the
Michelson interferometer, respectively, see
Sec.~\ref{sec:lasernoise} for their significance.
\label{fig:fields}}
\end{figure}

The interferometer configuration is shown in
Fig.~\ref{fig:ponderomotive_layout}, and in Fig.~\ref{fig:fields}
we show fields propagating in the interferometer as well as modes
of motion of the mirrors. The interferometer is similar to that
used in GW detection: a Michelson interferometer with Fabry-Perot
cavities in each arm. All the mirrors of the interferometer are
suspended as pendulums. Power-recycling~\cite{Drever} is optional
and is not included here. The configuration shown has a few
unusual features compared with a conventional interferometer,
however. First, the end mirrors of the arm cavities are a common
suspended object, coated with a high-reflectivity coating on both
surfaces and assumed to have an opaque substrate. Second, this
cavity end mirror object is very light, with a typical mass of
1~g, and is suspended as a pendulum with resonant frequency of
about 1~Hz. All remaining optics are assumed to have a mass of
250~g, and are also suspended as pendulums with a resonant
frequency of 1~Hz. Third, the cavities are detuned from resonance.

Testing the simulation with this somewhat unconventional
interferometer configuration served two purposes: (i) It is the
baseline design for an experiment to generate squeezed states of
the electromagnetic field, produced with
radiation-pressure-induced optical forces in an interferometer
with low-mass mirror oscillators and high stored power
\cite{ponderomotive}; and (ii) the shared end mirror gives rise to
unexpected dynamical effects that prove interesting and
instructive to explore, and are relevant to other high-power
interferometers, such as Advanced LIGO~\cite{pfspie}. We note that
the shared end mirror has advantages in terms of mechanical
stability and control system design, but the desired
radiation-pressure effects can be realized by a configuration with
two independent end mirrors as well.

%------------ Table of parameters -----------------------
%\begin{widetext}
\begin{table}[ht]
\begin{tabular}{lclll}
\hline
Parameter & \hspace{-0.5cm} Symbol &  Value & Units\\
\hline \hline
%Light frequency & $\omega$ & $1.8 \times 10^{15} {\rm ~s}^{-1}$ \\
Light wavelength & $\lambda_{0}$ & $1064$&nm\\
End mirror mass & $m$ & $1$&g\\
Input mirror mass & $M$ & $0.25$&kg\\
Input mirror transmission & $T_i$ & $4 \times 10^{-4}$ & --\\
Arm cavity finesse & $\cal F$ & $1.6 \times 10^{4}$ & --\\
Loss per bounce & -- & $5 \times 10^{-6}$ & --\\
Arm cavity detuning & $\phi$ & $10^{-5}$& $\lambda_{0}$\\
Input power & $I_{0}$ & $1$ & W\\
\hline
BS reflectivity asymmetry & $\Delta_{\mbox{\tiny BS}}$ & $0.01$ & --\\
Michelson phase imbalance & $\Delta \alpha_{\mbox{\tiny M}}$ & & \\
Michelson loss imbalance & $\Delta \epsilon_{\mbox{\tiny M}}$ & & \\
Input mirror mismatch & $\Delta_{T}$ & $5 \times 10^{-6}$ & --\\
Detuning mismatch & $\Delta_{\phi}$ & $10^{-7}$ & $\lambda_{0}$\\
Arm cavity loss mismatch & $\Delta_{\epsilon}$ & $2 \times 10^{-6}$ & --\\
\hline \hline
\end{tabular}
\caption{Select interferometer parameters and their nominal
values. \label{table:parameters} }
\end{table}
%\end{widetext}

{\renewcommand{\arraystretch}{1}
\begin{table}[bth]
\centerline{
\begin{tabular}{ccc}
\hline\hline
$\epsilon$ & bandwidth & $(T_i+T_e)c/(4L)$\\
$\epsilon_{\rm L}$ & bandwidth due to loss & $T_e c/(4L)$\\
$-\lambda$ & resonant frequency & $\phi c/L$\\
$\alpha$ & characteristic quadrature rotation angle &
$\arctan(\lambda/\epsilon)$
\\
\hline\hline
\end{tabular}}
\caption{Quantities associated with the detuned arm cavities
.\label{tab:basic}}
\end{table}
}

\subsection{Ideal optical springs}
\label{sec:optsprings}

In this section we study analytically a crucial component of the
interferometer design: the optical spring effect, especially in
the case of two identical detuned cavities with a common end
mirror. The input-output relation of this system can be obtained
by carrying out our generic procedure analytically. In doing so,
we extend previous results in Refs.~\cite{BC2,BC5} to include two
new features. First, we consider motions of all {\it three}
mirrors, with mass of the input mirrors different from that of the
common end mirror. Second, in our system the carrier phases
incident on mirrors are different; under such a circumstance,
formulas developed in Sec.~\ref{sect:radpres} are non-trivial
extensions to existing ones.

In order to make results intuitively understandable, we consider
only the ideal system, with the two input mirrors completely
identical, the common mirror perfectly reflective on both sides,
the two cavities having exactly the same lengths, the carrier
incident on both input mirrors having equal amplitude and phase,
and with a perfect beamsplitter. We also ignore the free pendulum
frequency, and consider the test masses to be free.  Similar to
previous studies, we assume a high-finesse cavity and ignore the
interaction between the motion of the input mirror and the carrier
light outside the cavity. We retain terms only to the leading
order in $\epsilon L/c$, $\lambda L/c$ and $\Omega L/c$, where $L$
is the cavity length, $c$ is the speed of light, $\Omega$ is the
sideband frequency, and $(-\lambda-i\epsilon)$ is the complex
optical resonant frequency of the cavity with fixed mirrors
[$-\lambda$ denotes the resonant frequency and $\epsilon$ the
bandwidth, defined in Table~\ref{tab:basic}; and we ignore
end-mirror loss].

\subsubsection{Differential Mode}

With the above assumptions, the differential optical mode couples
only to differential modes of mirror motion: those with the two
input masses moving such that $x_A=-x_B\equiv\tinysub{x}{D}$, and
arbitrary $x_m$ [see Fig.~\ref{fig:fields}]; such modes form a
two-dimensional subspace of all possible motions of the three
mirrors. In the ideal case, we only need to study this mode. The
differential input-output relation is given by
\begin{widetext}
\beq
\label{input-output:diff}
\left(\begin{array}{c} b_1 \\
b_2 \end{array}\right) = \frac{1}{\tinysub{\mathcal{M}}{D}}\mathbf{R}_{\alpha}\left[
\tinysub{\mathbf{C}}{D}\mathbf{R}_{-\alpha} \left(\begin{array}{c} a_1 \\ a_2
\end{array}\right) + \tinysub{\mathbf{s}}{D} \left[x_m^{(0)}+\tinysub{x}{D}^{(0)}\right]
\right]\,, \eeq
with
\bea
\label{input-output:C:diff}
\tinysub{\mathbf{C}}{D}
=\left[
\begin{array}{cc}
-(\Omega^2-\lambda^2+\epsilon^2)\Omega^2-\lambda\tinysub{\iota}{D}
&
2\epsilon\lambda\Omega^2 \\
-2\epsilon\lambda\Omega^2 + 2\epsilon \tinysub{\iota}{D} &
-(\Omega^2-\lambda^2+\epsilon^2)\Omega^2- \lambda
\tinysub{\iota}{D}
\end{array}
\right] \,,\qquad
\tinysub{\mathbf{s}}{D}=\frac{2\sqrt{{\epsilon \tinysub{\iota}{D} \Omega^2 }}}{L h_{\rm SQL}^{\mbox{\tiny D}}}
 \left(\begin{array}{c} \lambda \\ -\epsilon+
i\Omega\end{array}\right)\,,
\eea
\end{widetext}
and \beq \label{calM:diff}
\tinysub{\mathcal{M}}{D}=\Omega^2\left[(\Omega+i\epsilon)^2-\lambda^2\right]+\lambda\tinysub{\iota}{D}\,.
\eeq Here $x_m^{(0)}$ is the motion of a free end mirror with the
same mass, $\tinysub{x}{D}^{(0)}$ is the free differential motion
of the input mirrors ($x_A^{(0)}=-x_B^{(0)}=
\tinysub{x}{D}^{(0)}$); $\alpha=\arctan(\lambda/\epsilon)$ is the
carrier phase at the end mirror. The carrier incident on the input
mirrors has phase 0, the carrier inside the cavity, leaving the
input mirror has phase $\alpha-\phi$, while the carrier inside the
cavity entering the input mirror has phase $\alpha+\phi$.  The
quantity $h_{\rm SQL}^{\mbox{\tiny D}}$ is the free-mass Standard
Quantum Limit associated with the differential mode, given by \beq
\label{sql:D} h_{\rm SQL}^{\mbox{\tiny
D}}=\sqrt{\frac{2\hbar}{\tinysub{\mu}{D}\Omega^2L^2}}\,,\qquad
\tinysub{\mu}{D} \equiv  2 m M/(m+2M)\,. \eeq The quantity
$\tinysub{\iota}{D}$, defined by \beq \label{eq:iota:D}
\tinysub{\iota}{D}=\frac{8\omega_0 I_c }{\tinysub{\mu}{D}L c}\,,
\eeq measures the strength of optomechanical coupling [notice the
dependence on carrier intensity $I_c$ and the inverse dependence
on the effective mass of the differential mode mechanical
oscillator $\tinysub{\mu}{D}$]. Roots of
$\tinysub{\mathcal{M}}{D}$ are the (complex) resonant frequencies
of the coupled optomechanical system. From $\tinysub{\iota}{D}$ we
define a characteristic frequency, \beq \label{eq:Theta}
\tinysub{\Theta}{D}\equiv
\sqrt{\tinysub{\iota}{D}\lambda/(\epsilon^2+\lambda^2)}\,. \eeq
For systems with $\tinysub{\Theta}{D} \ll \epsilon$, the two
resonances are well separated, and are given approximately by $\pm
\tinysub{\Theta}{D}$ [mechanical frequency due to optical spring]
and $(\pm \lambda - i\epsilon)$ [optical resonant frequency],
respectively --- this is indeed the regime in which we construct
our experiment.

The differential optical mode couples to a two-dimensional
subspace of all possible motions of the three mirrors.  It is
instructive to look at the motion of separate mirrors, in the
regime of $\Omega \ll \epsilon$, i.e. for sideband frequencies
$\Omega$ well within the linewidth of the cavities: \bea
\label{xmxM} \left[
\begin{array}{c}
x_m \\
\tinysub{x}{D}
\end{array}
\right]
\!\!&=&\!\!
\frac{1}{\tinysub{\Theta}{D}^2-\Omega^2} \nonumber \\
&&\!\!\left[
\begin{array}{cc}
\displaystyle \frac{\tinysub{\Theta}{D}^2}{\Lambda^2+1}-\Omega^2 & \displaystyle -\frac{\Lambda^2\tinysub{\Theta}{D}^2}{\Lambda^2+1} \\
\displaystyle -\frac{\tinysub{\Theta}{D}^2}{\Lambda^2+1} & \displaystyle
\frac{\Lambda^2\tinysub{\Theta}{D}^2}{\Lambda^2+1}-\Omega^2
\end{array}
\right] \left[
\begin{array}{c}
x_m^{(0)} \\
\tinysub{x}{D}^{(0)}
\end{array}
\right] .\quad\quad \eea Here we have defined $\Lambda ^2 \equiv
2M/m$. From Eq.~\eqref{xmxM}, we conclude immediately that \beq
\label{totallength} x_m+ \tinysub{x}{D} =
-\frac{\Omega^2}{\tinysub{\Theta}{D}^2-\Omega^2}\left[x_m^{(0)}+
\tinysub{x}{D}^{(0)}\right] \,. \eeq This change in response is
exactly what happens when a free test particle is connected to a
spring with mechanical resonant frequency $\tinysub{\Theta}{D}$.
Equation \eqref{totallength} reveals a crucial advantage of the
optical spring --- that the response of the cavity length to
external disturbances (driven by seismic and/or thermal forces,
e.g.) is greatly suppressed from the corresponding value for
free-mass systems. Theoretically, this suppression is present even
when a mechanical spring is used. However, mechanical springs
introduce thermal noise, usually of much higher magnitude due to
the intrinsic mechanical loss~\cite{Rigidity,BC5}.

It is interesting to notice that the suppression of total cavity
length fluctuations is achieved collectively by the end mirror and
the input mirror. As we see from Eq.~\eqref{xmxM}, [in the case of
large $\Lambda$], the motion of the end mirrors $\tinysub{x}{m}$
is suppressed from its free mass value by the factor in
Eq.~\eqref{totallength}, while the motion of the input mirrors
$\tinysub{x}{D}$ is not influenced by the spring, since it is
relatively massive. Fortunately, through the $(1,2)$ component of
the matrix on the RHS of Eq.~\eqref{xmxM}, this motion of the
input mirror is imposed onto the end mirror with opposite sign,
again suppressing the total cavity length fluctuations.

Now let us restrict ourselves to the regime of
$\Omega<\tinysub{\Theta}{D}<\epsilon$, and study the quantum
fluctuations and classical component of the output field (due to
classical disturbances to the mirrors). As we shall see shortly,
this regime has two crucial features: (i) the response of the
output field to $x_m^{(0)}+\tinysub{x}{D}^{(0)}$, and thus length
fluctuations due to seismic and thermal noise, are greatly
suppressed by the optical spring and (ii) the output squeezed
state is frequency-independent.

For quantum fluctuations, we have \beq \frac{\tinysub{\mathbf{C}}{D}}{\tinysub{\mathcal{M}}{D}}
\rightarrow \left[
\begin{array}{cc}
-1 & 0 \\
2 \epsilon/\lambda  & -1
\end{array}
\right]\,, \eeq which is frequency-independent. It is
straightforward to derive that the quantum noise spectrum in the
$b_\zeta\equiv b_1\cos\zeta+b_2\sin\zeta$ quadrature
[Cf.~Eq.~\eqref{eq:qnoise}]: \beq S_\zeta \rightarrow
1+\frac{2\epsilon^2}{\lambda^2}
-2\sqrt{\frac{\epsilon^2}{\lambda^2}+\frac{\epsilon^4}{\lambda^4}}
\cos(2\zeta-3\alpha) \,. \eeq In particular, terms in
$\epsilon/\lambda$ are associated with squeezing, where the
constant power squeeze factor $e^{2q}$ ($q>0$) is given by \beq
\label{sinhq} \sinh q = |\epsilon/\lambda|\,. \eeq The minimum
noise spectral density ($S_ \zeta=e^{-2q}$) is reached at $\zeta=3
\alpha/2$, while at $\zeta=\alpha$ and $2\alpha$ the noise
spectrum is equal to the vacuum level ($S_\zeta=1$). Values of
$\epsilon/\lambda$ corresponding to several power squeeze factors
are listed in Table~\ref{tab:sinhq}. As shown, $\epsilon$ and
$\lambda$ will not differ by a factor of more than $\sim 2$, for
typically desired squeeze factors.

Now for the classical component, given by the second term in
Eq.~\eqref{input-output:diff}, we have \beq \label{smsimp}
\frac{1}{\tinysub{\mathcal{M}}{D}}\mathbf{R}_\alpha\tinysub{\mathbf{s}}{D}
\rightarrow \frac{2}{L h_{\rm SQL}^{\mbox{\tiny D}} }
\sqrt{\frac{{\Omega}^2}{\tinysub{\Theta}{D}^2}
\frac{\epsilon}{\lambda}} \left[
\begin{array}{c}
\sin2\alpha \\ -\cos2\alpha
\end{array}
\right]\,. \eeq This means the entire signal due to differential
displacement $x_m^{(0)}+\tinysub{x}{D}^{(0)}$ is in the single
quadrature $\zeta=2\alpha+\pi/2$, and there is no
$x_m^{(0)}+\tinysub{x}{D}^{(0)}$ signal in the $\zeta=2\alpha$
quadrature. Interestingly, the quantum noise in this quadrature is
right at vacuum level. In addition, since
$\tinysub{h}{SQL}^{\mbox{\tiny D}} \propto 1/\Omega$, the response
of $b_\zeta$ to $x_m^{(0)}+\tinysub{x}{D}^{(0)}$ is proportional
to $\Omega^2$ at this regime -- therefore not only the motion, but
also the output field, has a suppressed response to thermal and
seismic noises. Note here that the suppression factor is
proportional to $\sqrt{I_c}$ (since $\tinysub{\theta}{D} \propto
\sqrt{\tinysub{\iota}{D}} \propto \sqrt{I_c}$) -- because motion
is suppressed by $I_c$, while the optical sensing of mirror motion
is enhanced by $\sqrt{I_c}$. Now suppose we introduce a noisy
force which induces a spectral density $S_x^N$ on a free mass,
then the output classical noise will be \beq S_\zeta^N =
4\frac{\Omega^2}{\Theta_D^2}\frac{\epsilon}{\lambda}\sin^2(\zeta-2\alpha)\frac{S_x}{L^2
(h_{\mbox{\tiny SQL}}^{\mbox{\tiny D}})^2}\,. \eeq At the minimum
quantum noise quadrature, $\zeta=3\alpha/2$, we have
\bea
S_{3\alpha/2}^N  &=& \frac{2\epsilon}{\lambda}
\left[
1-\frac{\epsilon}{\sqrt{\lambda^2+\epsilon^2}}\right] \frac{\Omega^2}{\Theta_D^2} \frac{S_x}{L^2
(h_{\mbox{\tiny SQL}}^{\mbox{\tiny D}})^2} \nonumber \\
& \le &0.6 \frac{\Omega^2}{\Theta_D^2}
 \frac{S_x}{L^2
(h_{\mbox{\tiny SQL}}^{\mbox{\tiny D}})^2}\,,
\eea
where the inequality is obtained by taking  maximum over all $\epsilon$ and $\lambda$.
We note that
because of the suppression factor $\Omega^2/\Theta^2$, the
classical noise $S^N_x$ can be much higher than the free-mass
Standard Quantum Limit while still allowing the interferometer to
generate squeezed vacuum!

%-------------
{\setlength{\tabcolsep}{5pt}
\begin{table}[ht]
\begin{tabular}{|c|cccc|}
\hline
 Squeeze Factor  (dB)
& 3 & 7 & 10 & 20 \\
\hline
 $\epsilon/\lambda$ & 0.58  & 1.13 & 1.42 & 2.12\\
 \hline
\end{tabular}
\caption{Relationship between power squeeze factor and
$\epsilon/\lambda$, see Eq.~\eqref{sinhq}. \label{tab:sinhq} }
\end{table}
}
%--------------

\subsubsection{Common Mode}

We now consider the common optical mode, which couples with motion
of the input mirrors corresponding to
$x_A=x_B\equiv\tinysub{x}{C}$. This mode is irrelevant to an ideal
interferometer with identical arms and perfect contrast. In
reality, however, the common mode {\it will} influence the output
via couplings induced by differences (mismatch) between the two
cavities, for example. Such effects can be quite important near
the common-mode optomechanical resonance.

The input-output relation of the common mode, similar to that of
the differential mode [cf.~Eq.~\eqref{input-output:diff}], is
given by: \beq \label{input-output:comm}
\left(\begin{array}{c} y_1 \\
y_2 \end{array}\right) = \frac{1}{\tinysub{\mathcal{M}}{C}}\mathbf{R}_{\alpha}\left[
\tinysub{\mathbf{C}}{C}\mathbf{R}_{-\alpha} \left(\begin{array}{c} z_1 \\ z_2
\end{array}\right) + \tinysub{\mathbf{s}}{C} \tinysub{x}{C}^{(0)}
\right]\,, \eeq with [cf.~Eq.~\eqref{input-output:C:diff}]
\begin{widetext}
\bea
\tinysub{\mathbf{C}}{C}
=\left[
\begin{array}{cc}
-(\Omega^2-\lambda^2+\epsilon^2)\Omega^2-\lambda\tinysub{\iota}{C}
&
2\epsilon\lambda\Omega^2 \\
-2\epsilon\lambda\Omega^2 + 2\epsilon \tinysub{\iota}{C} &
-(\Omega^2-\lambda^2+\epsilon^2)\Omega^2- \lambda
\tinysub{\iota}{C}
\end{array}
\right] \,,\qquad
\tinysub{\mathbf{s}}{C}=\frac{2\sqrt{{\epsilon \tinysub{\iota}{C} \Omega^2 }}}{L h_{\rm SQL}^{\mbox{\tiny C}}}
 \left(\begin{array}{c} \lambda \\ -\epsilon+
i\Omega\end{array}\right)\,,
\eea
\end{widetext}
and [cf. Eq.~\eqref{calM:diff}] \beq \label{calM:comm}
\tinysub{\mathcal{M}}{C}=\Omega^2\left[(\Omega+i\epsilon)^2-\lambda^2\right]+\lambda\tinysub{\iota}{C}\,.
\eeq $h_{\rm SQL}^{\mbox{\tiny C}}$, the SQL associated with the
common mode, is given by [cf.~Eq.~\eqref{sql:D}] \beq h_{\rm
SQL}^{\mbox{\tiny C}}
=\sqrt{\frac{2\hbar}{\tinysub{\mu}{D}\Omega^2L^2}}\,,\quad
\tinysub{\mu}{C}=2M\,. \eeq The quantity $\tinysub{\iota}{C}$ is
given by [cf.~Eq.~\eqref{eq:iota:D}] \bea
\tinysub{\iota}{C}=\frac{8\omega_0 I_c}{\tinysub{\mu}{C}Lc}\,.
\eea For the common mode, we have a optomechanical resonant
frequency of [cf.~Eq.~\eqref{eq:Theta}] \beq \label{eq:Theta:comm}
\tinysub{\Theta}{C}\equiv
\sqrt{\tinysub{\iota}{C}\lambda/(\epsilon^2+\lambda^2)}\,,\quad
{\mbox{if}}\; \tinysub{\Theta}{C} \ll \epsilon\,.\qquad \eeq This
frequency is in general much lower than its differential-mode
counterpart, with \beq \label{eq:resonance:ratio}
\frac{\tinysub{\Theta}{C}}{\tinysub{\Theta}{D}}
=\sqrt{\frac{\tinysub{\iota}{C}}{\tinysub{\iota}{D}}}
=\sqrt{\frac{\tinysub{\mu}{D}}{\tinysub{\mu}{C}}}
=\sqrt{\frac{m}{m+2M}}\,. \eeq

%%%%%%%%%%%%%
{\setlength{\tabcolsep}{10pt}
\renewcommand{\arraystretch}{2}
\begin{table}
\begin{tabular}{c|cc}
\hline\hline
$\Delta_{(k)}$ & $C_{(k)}$ & $\varphi_{(k)}^{\rm C}$ \\
\hline
$\displaystyle \frac{\Delta\epsilon}{\epsilon}$ & $\displaystyle -\frac{\epsilon\lambda}{\epsilon^2+\lambda^2}$ & $2\alpha+\pi/2$\\
$\displaystyle \frac{\Delta \subl{\epsilon}}{\epsilon}$ & $\displaystyle -\frac{\epsilon}{\sqrt{\epsilon^2+\lambda^2}}$ & $\alpha$ \\
$\displaystyle \frac{\Delta\lambda}{\lambda}$ & $\displaystyle \frac{\epsilon\lambda}{\epsilon^2+\lambda^2}$ & $2\alpha+\pi/2$ \\
$\displaystyle \subm{\Delta\alpha}$ & $1$ & $2\alpha+\pi/2$ \\
$\displaystyle \subbs{\Delta}$ & 0 & \\
$\displaystyle \subm{\Delta\epsilon}$ & $-\frac{1}{2}$ & $2\alpha$\\
\hline\hline
\end{tabular}
\caption{\label{tab:MMcarrier} Transfer function from carrier to
differential output [see Eq.~\eqref{outputC}], in the
leading-order approximation. The same coefficients apply to
phase-noise coupling, i.e., $N_{k}^{\rm P}=C_{(k)}$,
$\varphi_{k}^{\rm P}=\varphi_{(k)}^{\rm C}$, in the low-frequency
regime [see Eq.~\eqref{outputnoise}].}
\end{table}
}
%%%%%%%%%%%%%%%%%%%

%%%%%%%%%%%%%%
{\setlength{\tabcolsep}{10pt}
\renewcommand{\arraystretch}{2.5}
\begin{table*}
\begin{tabular}{c|ccc}
\hline\hline
$\Delta_{(k)}$ & $N_{(k)}^A$ & $N_{(k)}^A(\tinysub{\Theta}{C}\rightarrow 0)$& $\varphi_{(k)}^A$ \\
\hline\hline $\displaystyle \frac{\Delta\epsilon}{\epsilon}$ &
$\displaystyle \frac{\epsilon^2 \left[
\epsilon^2(\Omega^2+\tinysub{\Theta}{C}^2)^2 +4\lambda^2
\tinysub{\Theta}{C}^4 \right]^{1/2}}{\lambda
(\epsilon^2+\lambda^2)(\Omega^2-\tinysub{\Theta}{C}^2)}$  &
$\displaystyle \frac{\epsilon^3}{\lambda(\epsilon^2+\lambda^2)}$ &
$\displaystyle 2\alpha-\arctan\frac{2\lambda
\tinysub{\Theta}{C}^2
}{\epsilon(\Omega^2+\tinysub{\Theta}{C}^2)}$
\\
\hline $\displaystyle \frac{\Delta\epsilon_{\rm L}}{\epsilon}$ &
$\displaystyle
\frac{\epsilon^2}{\lambda\sqrt{\epsilon^2+\lambda^2}}$ &
$\displaystyle
\frac{\epsilon^2}{\lambda\sqrt{\epsilon^2+\lambda^2}}$ & $\alpha$
\\
\hline $\displaystyle \frac{\Delta\lambda}{\lambda}$ &
$\displaystyle \frac{\epsilon\left[(\lambda^2\Omega^2 -\epsilon^2
\tinysub{\Theta}{C}^2)^2+4 \epsilon^2\lambda^2\tinysub{\Theta}{C}^4\right]^{1/2}}{\lambda(\epsilon^2+\lambda^2)(\Omega^2-\tinysub{\Theta}{C}^2)}$ & $\displaystyle
\frac{\epsilon\lambda}{\epsilon^2+\lambda^2}$ & $\displaystyle
2\alpha+\arctan\frac{2\epsilon\lambda\tinysub{\Theta}{C}^2}{\lambda^2\Omega^2 - \epsilon^2 \tinysub{\Theta}{C}^2}$
\\
\hline $\Delta\alpha_{\mbox{\tiny M}}$ &$\displaystyle
-\frac{\left[\lambda^2(\Omega^2-\tinysub{\Theta}{C}^2)^2+4
\epsilon^2\tinysub{\Theta}{C}^4\right]^{1/2}}{\lambda(\Omega^2-\tinysub{\Theta}{C}^2)}$ & $-1$ &
$\displaystyle 2\alpha-\arctan\frac{2\epsilon\tinysub{\Theta}{C}^2}{\lambda(\Omega^2-\tinysub{\Theta}{C}^2)}$
\\
\hline $\Delta_{\mbox{\tiny BS}}$ & $\displaystyle \frac{2\epsilon
\Omega^2}{\lambda(\Omega^2-\tinysub{\Theta}{C}^2)}$ & $\displaystyle
\frac{2\epsilon}{\lambda}$ &$2\alpha$
\\
\hline $\Delta\,\epsilon_{\mbox{\tiny M}}$ & $\displaystyle
\frac{\left[
\left[(\epsilon^2+\lambda^2)\Omega^2-(2\epsilon^2+\lambda^2)\tinysub{\Theta}{C}^2 \right]^2+\epsilon^2\lambda^2\tinysub{\Theta}{C}^4 \right]^{1/2}
}{2\lambda\sqrt{\epsilon^2+\lambda^2}(\Omega^2-\tinysub{\Theta}{C}^2)}$
& $\displaystyle \frac{\sqrt{\epsilon^2+\lambda^2}}{2\lambda}$ &
$\displaystyle \alpha -\arctan\frac{\epsilon \lambda
\tinysub{\Theta}{C}^2}{(\epsilon^2+\lambda^2)\Omega^2-(2\epsilon^2+\lambda^2)
\tinysub{\Theta}{C}^2}$
\\
\hline\hline
\end{tabular}
\caption{\label{tab:laseramp} Laser amplitude noise coupling into the dark port, in the leading-order approximation and low-frequency regime [see Eq.~\eqref{outputnoise}].}
\end{table*}
}

\subsection{Laser coupling to the antisymmetric port due to mismatch}
\label{sec:lasernoise}

Mismatch between the optical parameters of the two arm cavities,
as well as imbalance in the beamsplitter reflection/transmission
ratio and imperfect contrast of the Michelson interferometer, can
couple the carrier and also the noise sidebands on the laser to
the differential detection port. For each arm, A and B, we denote
the true value of the $k$th quantity by its nominal value plus
contributions due to imperfections, i.e. \beq X_{(k)
{A,B}}=X_{(k)} \pm \frac{1}{2}\Delta X_{(k)}\,. \eeq Here the
index $k$ refers to the type of imperfection being considered. The
beamsplitter asymmetry is characterized by \beq \label{DBS}
\subbs{\Delta} = \subbs{t}^2-\subbs{r}^2\,. \eeq Michelson
imperfections can be characterized by the difference in the phase
shifts and losses when light travels from the beamsplitter to the
input mirrors of the two arms: \beq \label{AEM}
\tinysub{\alpha}{MA,B}=\subm{\alpha} \pm
\frac{1}{2}\tinysub{\Delta\alpha}{M}\,, \qquad
\tinysub{\epsilon}{MA,B}=\subm{\epsilon} \pm
\frac{1}{2}\tinysub{\Delta\epsilon}{M}\,. \eeq

In addition to $\subbs{\Delta}$, $\Delta\subm{\alpha}$ and
$\Delta\subm{\epsilon}$, which concern the beamsplitter, we
consider the following contributions to mismatch between the arms,
\bea
T_{i{\rm A,B}}&\equiv&T_i\pm \frac{1}{2}\Delta_T\,, \\
T_{e{\rm A,B}}&\equiv&T_e\pm \frac{1}{2}\Delta_\epsilon\,, \\
\phi_{\rm A,B}&\equiv&\phi\pm\frac{1}{2}\Delta_\phi\,, \eea that
is, mismatch between input mirror power transmissivities, end
mirror losses and cavity detuning, respectively. We replace these
with the following more convenient quantities: \beq \frac{\Delta
\epsilon}{\epsilon} = \frac{\Delta_T}{T_i+T_e}\,,\;\; \frac{\Delta
\tinysub{\epsilon}{L}}{\epsilon} =
\frac{\Delta_\epsilon}{T_i+T_e}\,, \;\;
\frac{\Delta\lambda}{\lambda}=\frac{\Delta_\phi}{\phi}\,. \eeq
[See Table~\ref{tab:basic} for definitions of $\epsilon$,
$\tinysub{\epsilon}{L}$ and $\lambda$.]

In the remainder of this section, we give the transfer functions
from the carrier light (DC), laser amplitude fluctuations and
laser phase fluctuations to the differential output port, to first
order in the mismatch (recall that ideally, in the absence of
imperfections, these common-mode inputs do not appear in the
differential output port). We keep our formulae to the leading
order in $\{ \Omega L/c,\epsilon L/c,\lambda L/c\}$, and ignore
the {\it averaged} losses, $\tinysub{\epsilon}{L}$ and
$\tinysub{\epsilon}{M}$ (but not $\tinysub{\Delta\epsilon}{L}$ and
$\tinysub{\Delta\epsilon}{M}$). We refer to this as the {\it
leading-order approximation}. Furthermore, in order to keep the
analytical results understandable, we work only in the regime of
$\{\Omega,\tinysub{\Theta}{C}\}  \ll \{\tinysub{\Theta}{D},
\lambda, \epsilon\}$, which we shall refer to as the {\it
low-frequency regime}.

Definitions and assumed values for $\tinysub{\Delta}{BS}$,
$\tinysub{\Delta \alpha}{M}$, $\tinysub{\Delta \epsilon}{M}$,
$\tinysub{\Delta}{T}$, $\Delta_{\phi}$, and $\Delta_{\epsilon}$
are given in Table~\ref{table:parameters}.

\subsubsection{The Carrier}

The transfer function from the carrier to the differential output
can be written as \beq \label{outputC} \sum_k \Delta_{(k)} C_{(k)}
\left(\begin{array}{c} \cos \varphi_{(k)}^{\rm C} \\  \sin
\varphi_{(k)}^{\rm C} \end{array}\right)\,, \eeq where definitions
of $\Delta_{(k)}$, values of $C_{(k)}$ and $\varphi_{(k)}^{\rm C}$
are listed in Table~\ref{tab:MMcarrier}, assuming the carrier at
the beamsplitter is in the first (amplitude) quadrature.

Contributions listed in Table~\ref{tab:MMcarrier} can all be
obtained from simple considerations. First, since each field that
interferes at the beamsplitter is scaled by one transmission and
one reflection coefficient factor, $\subbs{\Delta}$ does not
contribute to the output carrier light at the differential port.
Then, for all mismatches except the loss, one only has to notice
that when the arm cavities are lossless, carrier light with
amplitude $D$ and phase $\varphi=0$ returns to the beamsplitter
with amplitude reduced to $(1-\subm{\epsilon})$, and quadrature
rotated by $2\alpha+2\subm{\alpha}$. As a consequence, the
differential output port gets
$(D/2)(-\subm{\Delta\epsilon})=(-\subm{\Delta\epsilon}/2)D$ in the
$\varphi=2\alpha$ quadrature (factor of 2 due to the
beamsplitter), and \bea
&&(D/2)\Delta[2\alpha+2\subm{\alpha}] \nonumber \\
&=&
\left[\frac{\epsilon\lambda}{\epsilon^2+\lambda^2}\left(-\frac{\Delta\epsilon}{\epsilon}+\frac{\Delta\lambda}{\lambda}\right)+\subm{\Delta\alpha}\right]D
\eea in the orthogonal quadrature, $\varphi=2\alpha+\pi/2$. The
effect of the loss mismatch can be understood when we decompose
the (complex) reflectivity of the cavity into a sum of two
components: \bea
\frac{\sqrt{R_e}e^{2i\phi}-\sqrt{R_i}}{1-\sqrt{R_iR_e}e^{2i\phi}}&=&\frac{1+i\lambda/\epsilon}{1-i\lambda/\epsilon}-\frac{\subl{\epsilon}}{\epsilon}\frac{2}{1-i\lambda/\epsilon} \nonumber \\
&=&e^{2i\alpha} -
\frac{\subl{\epsilon}}{\epsilon}\frac{2\epsilon}{\sqrt{\epsilon^2+\lambda^2}}e^{i\alpha}
\eea Here we see that the loss $\subl{\epsilon}$ creates an output
at the $\varphi=\alpha$ quadrature, so an imbalance in loss
$\subl{\Delta\epsilon}$ will contribute \beq
\left(-\frac{\subl{\Delta\epsilon}}{\epsilon}
\frac{\epsilon}{\sqrt{\epsilon^2+\lambda^2}}\right)D \eeq in the
$\varphi=\alpha$ quadrature in the differential output port.

\subsubsection{Amplitude (Intensity) and Phase (frequency) Noise}

Under our simplifications, the laser amplitude noise
$z_1$ and phase noise $z_2$ couple to single (yet frequency
dependent) quadratures in the differential output port, as
parametrized by
\begin{eqnarray}
\label{outputnoise}
\sum_k \Delta_{(k)} &\Bigg[& N_{(k)} ^{\rm A}
\left(\begin{array}{c} -\sin\varphi_{(k)}^{\rm A} \\
\cos\varphi_{(k)}^{\rm A} \end{array}\right) z_1  \nonumber \\
&+& N_{(k)} ^{\rm
P} \left(\begin{array}{c} -\sin\varphi_{(k)}^{\rm P} \\
\cos\varphi_{(k)}^{\rm P} \end{array}\right) z_2\Bigg]\,.
\end{eqnarray}
Measurement of the output quadrature $b_{\zeta}\equiv b_1
\cos\zeta +b_2\sin\zeta$ will include the laser noise:
\begin{equation}
\sum_{(k)} \Delta_{(k)} \left[ N_{(k)}^{\rm A}
z_1\sin(\varphi_{(k)}^{\rm A}-\zeta) +
 N_{(k)}^{\rm P} z_2\sin(\varphi_{(k)}^{\rm P}-\zeta) \right]\,.
\end{equation}
In particular, the output quadrature $\zeta=\varphi_{(k)}^{\rm A
(P)}$ is not sensitive to the $k$-th contribution of laser
amplitude (phase) noise [note that we have switched the notation
for $\varphi$ from that of Eq.~\eqref{outputC}].

As it also turns out, in the leading-order approximation and the
low-frequency regime, $\left[N_{(k)}^{\rm P},\varphi_{(k)}^{\rm
P}\right]=\left[C_{(k)},\varphi_{(k)}\right]$. Considering the
different ways $\varphi$ appears in Eqs.~\eqref{outputC} and
\eqref{outputnoise}, this means the phase noise coupled to the
differential output port remains orthogonal to the carrier. This
can be argued for easily: since phase modulations on the carrier
do not drive mirror motion, the propagation of phase noise is not
affected by the optical spring. Amplitude modulations, on the
other hand, do drive mirror motion and therefore should couple to
the differential port in a dramatically different way.  We
tabulate the quantities $N_{(k)}^{\rm A}$ and $\varphi_{(k)}^{\rm
A}$ in Tab.~\ref{tab:laseramp}, from which we can see that the
amplitude-noise coupling has features around the common-mode
optical-spring resonant frequency, $\tinysub{\Theta}{C}$.

\subsubsection{Evading Laser Noise by Artificial Asymmetry}
\label{sec:laser_evasion}

For realistically achievable symmetry between the two arms, laser
noises turn out to be the dominant noise source to our squeezer.
Here we discuss a novel way of mitigating laser noise coupling by
introducing {\it artificial asymmetries}. According to the
approximate results (in the leading-order approximation and
low-frequency regime) obtained in the previous section, both
amplitude and phase noise emerge from single quadratures (as
vector sums of contributions from different mechanisms). We can,
therefore, eliminate the laser noise totally, up to this order, if
we make both of them emerge from the same quadrature
$\zeta+\pi/2$, and make sure that the orthogonal quadrature,
$\zeta$, has a sub-vacuum noise spectrum. At our disposal are two
asymmetries that we can adjust manually:
$\Delta\tinysub{\alpha}{M}$ and $\Delta \tinysub{\epsilon}{M}$.

At any given sideband frequency $\Omega$, for a generic set of
other asymmetries, it is always possible to make both laser noise
sources emerge at the $\zeta+\pi/2$ quadrature (and, therefore, to
vanish at the $\zeta$ quadrature), by adjusting
$\Delta\tinysub{\alpha}{M}$ and $\Delta \tinysub{\epsilon}{M}$, if
the following {\it non-degeneracy condition} is satisfied: \bea
&&\Delta_{\rm laser}(\Omega,\zeta) \nonumber \\
&\equiv& \det\left[
\begin{array}{cc}
\sin(\varphi_{\tinysub{\alpha}{M}}^{\rm A}-\zeta) N_{\Delta\tinysub{\alpha}{M}}^{\rm A}
&
\sin(\varphi_{\Delta\tinysub{\epsilon}{M}}^{\rm A}-\zeta) N_{\Delta\tinysub{\epsilon}{M}}^{\rm A}
\\
\sin(\varphi_{\tinysub{\alpha}{M}}^{\rm P}-\zeta) N_{\Delta\tinysub{\alpha}{M}}^{\rm P}
&
\sin(\varphi_{\Delta\tinysub{\epsilon}{M}}^{\rm P}-\zeta) N_{\Delta\tinysub{\epsilon}{M}}^{\rm P}
\end{array}
\right]\nonumber \\
&\neq& 0\,.
\eea
[See Eq.~\eqref{outputnoise}.]

According to Tables~\ref{tab:MMcarrier} and \ref{tab:laseramp},
laser phase noise emerges in a frequency-independent quadrature,
but the amplitude noise does not. This means the elimination of
laser noise must be frequency-dependent, and we can only choose
one particular frequency for perfect laser noise evasion. However,
if $\Omega \gg \tinysub{\Theta}{C}$ is also satisfied, then the
frequency-dependence goes away. We consider this special case, and
choose a detection quadrature of $\zeta=3\alpha/2$, i.e., the one
with minimum quantum noise. From Tabs.~\ref{tab:MMcarrier} and
\ref{tab:laseramp}, we get \beq \left.\Delta_{\rm
laser}\left(\Omega,\frac{3}{2}\alpha\right)\right|_{\rm
\tinysub{\Theta}{C}\rightarrow 0} =
-\frac{\epsilon}{4\sqrt{\epsilon^2+\lambda^2}}\neq 0\,. \eeq Since
the carrier always emerges $\pi/2$ away from the phase noise, it
emerges in exactly the same quadrature we propose to detect. In
this way, the laser-noise-evading squeezer always produces {\it
squeezed light with amplitude squeezing}.

Finally, we note that, due to possible higher-order corrections,
laser noise evasion may not be as perfect as predicted by our
first-order approximation, even at a single frequency. The amount
of {\it residual} laser noise, as well as the exact level of the
deliberate asymmetries we introduce, must be given by a more
accurate calculation.

%---------------------------------------------------------
\begin{figure*}[t]
\begin{center}
\begin{tabular}{cc}
\includegraphics[width=0.45\textwidth]{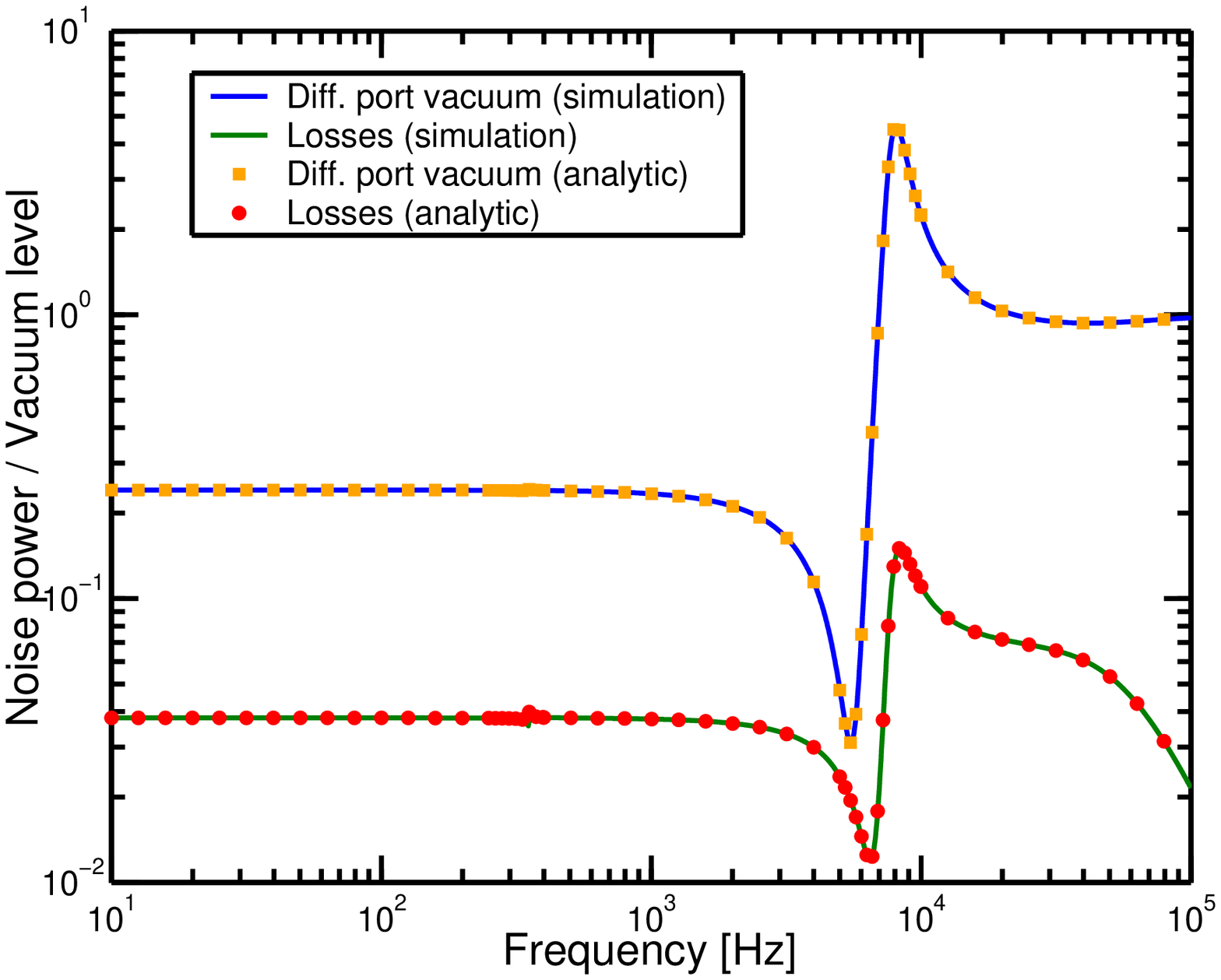}
\hspace{0.025\textwidth} & \hspace{0.025\textwidth}
\includegraphics[width=0.45\textwidth]{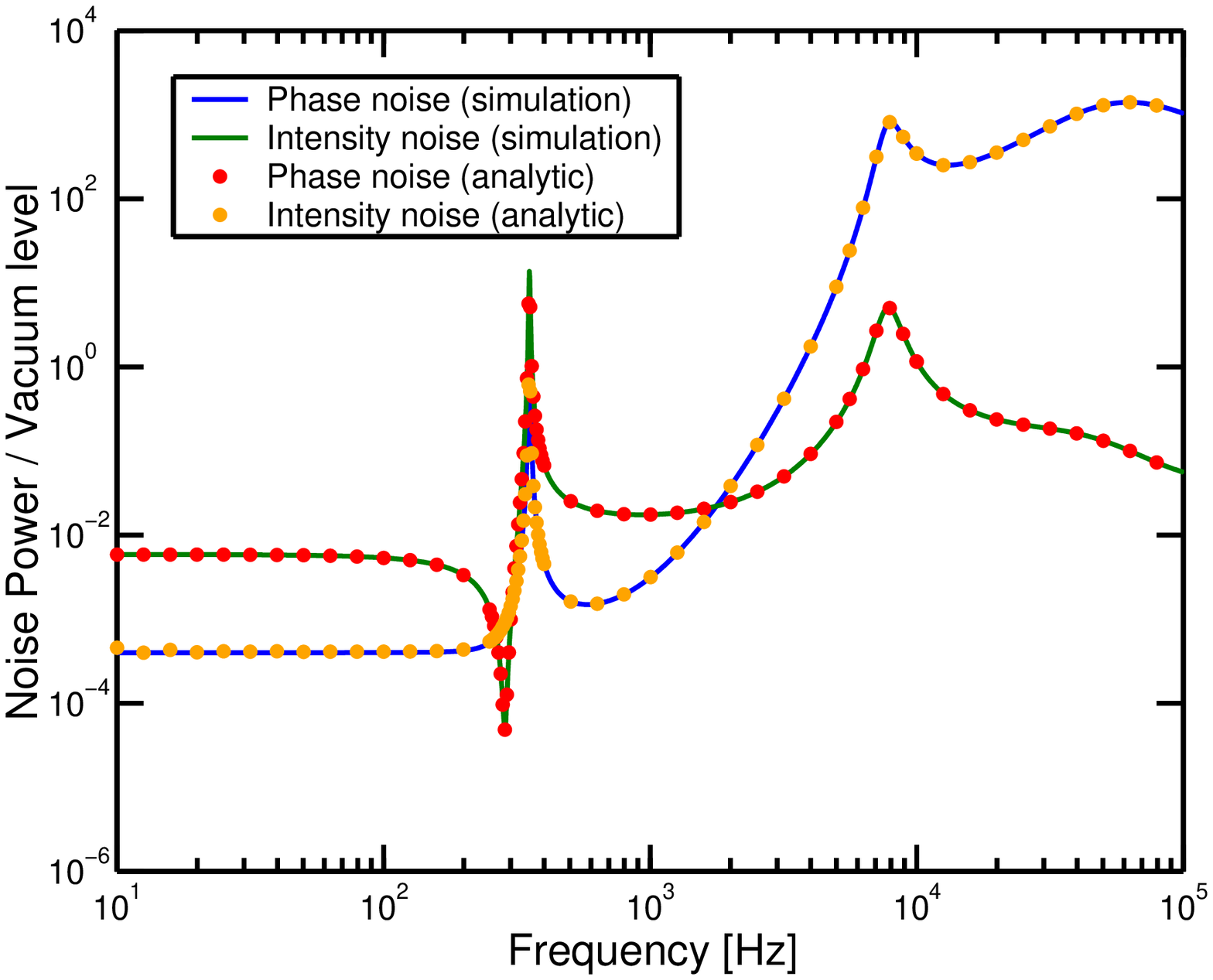}
\end{tabular}
\end{center}
\caption{Spectra of noise power at the output port of the
ponderomotive interferometer, normalized to the vacuum noise
level. The noise power is dimensionless, as compared to vacuum; a
pure vacuum (or shot noise) corresponds to unity. The lines are
results of the simulation code, while the data points are values
calculated from the corresponding analytic calculations. In the
left panel the vacuum noise level of the light exiting the
antisymmetric port is shown. The solid (blue) curve shows the
vacuum noise due to the (unsqueezed) vacuum fluctuations that
enter via the antisymmetric port of the interferometer; the dashed
(green) plot represents the noise due to the vacuum fluctuations
that enter via other optical losses in the system. At all
frequencies where the vacuum noise power is below unity, the
vacuum modes exiting the interferometer are squeezed due to
radiation-pressure effects. For the squeezing to be useful, all
noise couplings must yield a lower noise power than the squeezed
vacuum. In the right panel we show the coupling of laser frequency
(solid, blue) and laser amplitude (dashed, green) noise fields to
the output port, as calculated by the simulation code. Noise
levels of $10^{-4}\,\mathrm{Hz/Hz}^{1/2}$ for frequency noise and
$10^{-8}\,\mathrm{Hz}^{-1/2}$ for amplitude noise are assumed at
the input to the interferometer; all other parameters are listed
in Table \ref{table:parameters}. \label{fig:ponderomotive_noise}}
\end{figure*}
%-------------------------------------------------------------------

\subsection{Comparison between analytical calculations and numerical simulations}

In Table~\ref{table:parameters}, we list the parameters used in
modelling our interferometer. An important feature of the
numerical code is that it can handle imperfections in the optics
quite naturally, while for analytical techniques the solution
becomes complicated rather dramatically when more ingredients are
added. To fully test this feature, we constructed a test case with
realistic imperfections. The imperfections included were those
mentioned in Sec.~\ref{sec:lasernoise}. Using the parameters
listed in Table~\ref{table:parameters}, we calculate the noise at
the differential port due to quantum fluctuations entering from
this port and from lossy mirrors, as well as laser amplitude and
phase fluctuations entering from the symmetric port.

In Fig.~\ref{fig:ponderomotive_noise}, we show the calculated
noise levels from numerical simulations in curves, while those
from the analytical treatment are shown as solid points. The
agreement between the two sets of calculations is reassuring. Now
we discuss these noise spectrum in more details. In the left panel
of the Fig.~\ref{fig:ponderomotive_noise}, we plot noises due to
vacuum fluctuations entering from the dark port (light curve and
points), and due to vacuum fluctuations entering from mirror
losses (dark curve and points). In both results, there is a rather
dramatic resonant feature around the differential-mode
optical-spring resonant frequency, at $\tinysub{\Theta}{D}\approx
8\,$kHz, as can be expected from Sec.~\ref{sec:optsprings}.  The
rather weak but still noticeable feature around the common-mode
optical-spring resonant frequency $\tinysub{\Theta}{C}\approx
360\,$Hz is solely due to optical parameter mismatch. In the right
panel, we show laser amplitude (light curve and dots) and phase
(dark curve and points) noises; we have introduced artificial
asymmetries $\tinysub{\alpha}{M}$ and $\tinysub{\epsilon}{M}$,
with values obtained empirically using the numerical simulation
code, such that both laser noise sources are evaded to a roughly
maximal extent at 1~kHz. For this reason, contributions to the
results shown here are largely higher order, and we cannot hope to
explain them using results obtained in Sec.~\ref{sec:lasernoise}.
Here we do observe dramatic features around both the
differential-mode and the common-mode optical-spring resonances.

Results in Fig.~\ref{fig:ponderomotive_noise} are also of
great significance for a practical reason: they show that the vacuum
modes exiting the interferometer are squeezed by a large factor
even in the presence of realistic estimates for optical losses
(left panel) and laser amplitude and phase noise (right panel).

%%%%%%%%%%%%%%%%%%%%%

\section{Summary and Conclusions}
\label{sect:conclusions}

The main purpose of this work was to develop a mathematical
framework for the simulation of quantum fields in a complex
interferometer that includes radiation pressure effects.  We work
in the linear regime around the operation point of this
interferometer; in this regime, after adopting the Heisenberg
picture of quantum mechanics, the quantum equations of motion
(Heisenberg operators) of observables are identical to classical
ones.

During the development of this framework, we augmented previous
treatments of mirrors (and beamsplitters) by allowing the carrier
phases at the four (eight) ports to be different. This extension
gives rise to the optical spring effect even without detuned
optical cavities.

Based on this mathematical framework, we developed a simulation
code that can allow arbitrary optical topologies, and applied it
to a specific example of the interferometer shown in
Fig.~\ref{fig:ponderomotive_layout}. This interferometer was shown
to be capable of squeezing the vacuum modes that enter -- and
subsequently exit -- the differential port of the beamsplitter. We
introduced optical spring effects by detuning the arm cavities as
a means of mitigating the detrimental effects of thermal noise. We
study not only the quantum noise, but also laser noise couplings
from the symmetric (input or bright) port to the output
(antisymmetric or dark) port. Good agreement was found between
numerical results given by this code and analytical ones derived
independently. This agreement makes us confident that the
simulation is working correctly for this rather complex
interferometer.

During our study of the laser noise couplings, we found a novel
method of evading the laser noise by introducing artificial but
controlled asymmetries. This is crucial for the practical
implementation of this interferometer, and is likely to find
applications in many other experiments.

Our simulation code is now being used in the detailed optical
design of the Advanced LIGO interferometer. We also envisage the
following extensions to the code in the near future:
\begin{itemize}
\item Allowing multiple carrier or rf sidebands, which may be relevant
to the modeling of squeezing experiments that use nonlinear
optical media, e.g., crystals, as well as the modeling of error
signals for control systems.
\item Incorporating the modeling of servo loops. Here we may rely on the
input from quantum control theory as to whether and how
realistically a Heisenberg treatment can describe a
electro-optical feedback system.
\item Allowing nonlinear media or other elements with ``custom'' dispersion relations.
\end{itemize}

\acknowledgments We thank our colleagues at the LIGO Laboratory,
especially Keisuke Goda and David Ottaway, for stimulating
discussions. We gratefully acknowledge support from National
Science Foundation grants PHY-0107417, PHY-0300345 and (for Y.C.)
PHY-0099568. Y.C.'s research was also supported by the David and Barbara Groce Fund at the San Diego Foundation, as well as the Sofja Kovalevskaja Programme
(funded by the German Federal Ministry of Education and Research). Y.C.~thanks the MIT LIGO Laboratory for support
and hospitality during his stay. \appendix

\section{Two-photon quantum optical formalism}
\label{app:quad}

We use the two-photon formalism developed by Caves and
Schumaker~\cite{CavesSchumaker,SchumakerCaves} to describe GW
interferometers with significant radiation-pressure effects. In
this formalism, any quasi-monochromatic optical field $A$ with
frequency near the carrier frequency $\omega$ is written as \bea
E(t) &=& E_1(t)\,\cos(\omega \, t) + E_2(t)\,\sin(\omega \,t) \nonumber \\
&=& \left[\begin{array}{cc} \cos\omega t & \sin\omega t
\end{array}\right] \left[\begin{array}{c}E_1(t) \\ E_2(t)
\end{array}\right]\,, \eea where $E_1(t)$ and $E_2(t)$ are called
{\it quadrature fields}, which vary at timescales much longer than
that of the optical oscillation, $1/\omega$. The quadrature
formalism replaces $E(t)$ by \beq \mathbf{E}  =
\left[\begin{array}{c}E_1(t) \\ E_2(t) \end{array}\right]\,. \eeq
The DC components of $E_{1,2}(t)$ can be regarded as monochromatic
carrier light. In particular, carrier light with amplitude $D
e^{i\varphi}$ is represented as \beq D e^{i\varphi}
\Leftrightarrow \left( D e^{i\varphi} \right) e^{-i\omega t}
\Leftrightarrow D \left(\begin{array}{c} \cos\varphi \\ \sin
\varphi\end{array}\right)\,. \eeq AC components of $E_{1,2}(t)$,
which we denote by $A_{1,2}(t)$,  are called {\it sideband
fields}, which are usually more convenient to study once
transformed into the frequency domain,

\beq \tilde{A}_{1,2}(\Omega) = \int_{-\infty}^{+\infty} A_{1,2}(t)
e^{i\Omega t} dt\,. \eeq

In quantum two-photon optics, it is convenient to use a particular
normalization for sideband fields:
\bea
&&A_{1,2}(t)\nonumber \\
=&&  \sqrt{\frac{4\pi\hbar\omega}{\mathcal{A}c}}
\int_0^{+\infty}\frac{d\Omega}{2\pi}
\left[a_{1,2}(\Omega)e^{-i\Omega t}+H.c.\right]\,. \eea In this
way, we have a convenient set of commutation relations (for
$\Omega \ll \omega$)~\cite{CavesSchumaker,SchumakerCaves}:
\begin{subequations}
\beq
[a_1,a_1']=[a_2,a_2']=[a_1,a_1'^\dagger]=[a_2,a_2'^\dagger]=0\,,
\eeq
\beq
[a_1,a_2'^\dagger]=-[a_2,a_1'^\dagger]=2\pi i\delta(\Omega-\Omega')\,.
\eeq
\end{subequations}
Here we have denoted $a_{1,2}\equiv a_{1,2}(\Omega)$, $a_{1,2}' \equiv a_{1,2}(\Omega')$.

%-------------------------------------------------------------------
%%% bibliography

%%%%%%%%%%%%%%%%%%%%%%%

\end{document}